\documentclass[10pt, a4paper]{article}
\usepackage[utf8]{inputenc}
\usepackage[english]{babel}
\usepackage{csquotes,xpatch}
\usepackage[top=3cm, footskip=1cm, left=1.5cm, right=1.5cm]{geometry}
\usepackage{authblk}
\usepackage{amsfonts, amssymb, amsmath}
\usepackage{xfrac}
\usepackage{textcomp,marvosym}
\usepackage{nameref,hyperref}
\usepackage[switch]{lineno}
\usepackage{gensymb}
\usepackage{microtype}
	\DisableLigatures[f]{encoding = *, family = * }
\usepackage{mathtools}
\usepackage{array}
\usepackage{float}
\usepackage{setspace}
\usepackage[hang, aboveskip=5pt,labelfont=bf,labelsep=period,justification=justified,textfont=small]{caption}
\usepackage{graphicx}
	\graphicspath{{Figures/}} 
\usepackage[dvipsnames]{xcolor}
\usepackage{subcaption}
\usepackage{changepage}

\usepackage[maxnames=1, style=numeric, backend=bibtex, sorting=none, bibstyle=science]{biblatex}
\bibliography{mybib}
\usepackage{lastpage,fancyhdr,graphicx}
	\fancyheadoffset[LR]{1.00in}
	\fancyfootoffset[LR]{1.00in}
\usepackage{epstopdf}
\newcommand{\EQ}{\begin{equation}}
\newcommand{\EE}{\end{equation}}
\newcommand{\EQA}{\begin{eqnarray}}
\newcommand{\EEA}{\end{eqnarray}}

\newcommand{\nint}{n}

\newlength\savedwidth
\newcolumntype{+}{!{\vrule width 2pt}}

\author[1]{Roberto Morán-Tovar}
\author[2]{Henning Gruell}
\author[2]{Florian Klein}
\author[1]{Michael Lässig}
\affil[1]{Institute for Biological Physics, Universität zu Köln, Köln, Germany.}
\affil[2]{Laboratory of Experimental Immunology, Institute of Virology, Faculty of Medicine and University Hospital Cologne, Universität zu Köln, Köln , Germany.}

\title{Stochasticity of infectious outbreaks and consequences for optimal interventions}
\date{}

\begin{document}  

\twocolumn[ 
\maketitle
	\begin{@twocolumnfalse}
		\begin{abstract}
Global strategies to contain a pandemic, such as social distancing and protective measures, are designed to reduce the overall transmission rate between individuals. Despite such measures, essential institutions, including hospitals, schools, and food producing plants, remain focal points of local outbreaks. Here we develop a model for the stochastic infection dynamics that predicts the statistics of local outbreaks from observables of the underlying global epidemics. Specifically, we predict two key outbreak characteristics: the probability of proliferation from a first infection in the local community, and the {\rm establishment size}, which is the threshold size of local infection clusters where proliferation becomes likely. We derive these results using a contact network model of communities, and we show how the proliferation probability depends on the contact degree of the first infected individual. Based on this model, we suggest surveillance protocols by which individuals are tested proportionally to their degree in the contact network. We characterize the efficacy of contact-based protocols as a function of the epidemiological and the contact network parameters, and we show numerically that such protocols outperform random testing. 		\end{abstract}
	\end{@twocolumnfalse}

] 

\newpage

\section*{Introduction}
{\huge T}he current SARS-CoV-2 pandemic caused millions of deaths and severely affected global health and economy~\cite{BarOn2020}. Before population-wide immunity is achieved, non pharmaceutical interventions (NPIs) are required to suppress or mitigate the transmission of the infections~\cite{Karin2020, Cowling2020, Flaxman2020}. Some measures can easily be implemented and cause only a small burden for the population, such as the usage of protective masks or contact tracing strategies. Other measures, such as massive testing, lockdowns and social distancing, can be costly and of large impact for the population~\cite{Walker2020}. Core institutions such as hospitals, schools, and food supply and producing plants must continue functioning in order to satisfy the basic needs of communities~\cite{Black2020}. Under these circumstances, NPIs can be combined in order to avoid local outbreaks. On the one hand, protective measures are implemented to reduce the so-called initial effective reproductive number $R_0$, i.e., the average number of new infections caused by an infected individual during its infectious period, slowing down the initial spread of the virus. Additionally, given the high fraction of asymptomatic infected individuals reported for SARS-Cov-2~\cite{Li2020, Rivett2020}, surveillance protocols based on regular testing and contact tracing can be performed to detect and control the silent spread of the virus. However, RT-PCR tests can be costly to perform at high coverage even in small populations. Alternative testing protocols, including pool testing~\cite{Mutesa2021} and the use of rapid tests, have been developed to tackle this problem. All such testing strategies respond to the stochastic dynamics of  local outbreaks; this link is at the core of the present paper. 

In the first part, we discuss an epidemiological model for outbreaks that relates key local characteristics to observables of the underlying global epidemic. Starting from the well-known deterministic Susceptible-Infected-Recovered (SIR) model, we build a more realistic model by integration of three important features. First, the infection dynamics of highly contagious pathogens is often delayed by an incubation period, which is accounted for by extending the epidemiological dynamics to a so-called Susceptible-Exposed-Infected-Recovered (SEIR) model \cite{Bailey1975, keeling2008}. Second, the infection, incubation, and recovery processes are intrinsically stochastic across different individuals; that is, the epidemiological model becomes stochastic. Third, the infection risks of individuals are highly disperse because of their contact intensity; this generates an additional stochasticity the infection dynamics that can be captured by a contact network model for local communities~\cite{Barabasi1999, Newman2002}. To characterize such networks and the resulting stochastic dynamics, we study the degree distribution $\hat p_k$, i.e., the probability that a given individual interacts with $k$ other individuals in the network. This approach allows for treatable analytical calculations and captures broad network features accessible prior to an outbreak. It differs from disordered out-of-equilibrium models that exploit the unidirectionality of an epidemic propagating in a contact network and have been used to infer the origin of the outbreak from posterior information \cite{Lokhov2014, Altarelli2014, Lokhov2015}. 

We  use the fully stochastic model to describe statistical characteristics of local outbreaks that are important for monitoring and interventions. We compute the proliferation probability for outbreaks starting from a single infected individual, where proliferation is defined as the transition to deterministic growth; the complementary fate of an outbreak is extinction due to stochastic fluctuations. Our analysis extends previous results of epidemic growth in well-mixed populations~\cite{Allen2012, Allen2017, Tritch2018} and on contact networks~\cite{Newman2002, Meyers2005, PastorSatorras2015}. Specifically, we show how the local outbreak statistics can be estimated from global epidemiological parameters, including the growth rate and the average incubation and infection periods. Our analytical results are tested by individual-based stochastic simulations of outbreaks on random contact networks~\cite{Barabasi1999}. 

In the second part, we explore surveillance protocols based on random testing of a fraction of the total asymptomatic population that can be implemented in a straightforward way. Recently, different studies have considered strategies of random testing in large and well-mixed populations~\cite{Muller2020}, where the interaction network between individuals is unknown. Other studies have highlighted the importance of the underlying structure of interactions in the population for the implementation of efficient containment strategies~\cite{Eubank2004, Firth2020, Baker2021} and of optimal surveillance protocols to protect patients and healthcare workers~\cite{Rivett2020, Qiu2021}. In this work, we study the consequences of infection and contact stochasticity on the optimal implementation of surveillance protocols. We propose strategies to improve such protocols, in order to maximize the probability of detection of infected individuals and to minimize the time before detection and the expected epidemic size. We present a novel network-based protocol that outperforms a protocol based on uniform random sampling of the population. We perform simulations of the surveillance protocol and test its performance numerically. 

\section*{Epidemiology of local outbreaks} 
\subsection*{Deterministic epidemiological model}

\begin{figure}[h!]
\vspace*{-3ex}
\centering
\begin{subfigure}{.8\columnwidth}
\includegraphics[width = \textwidth]{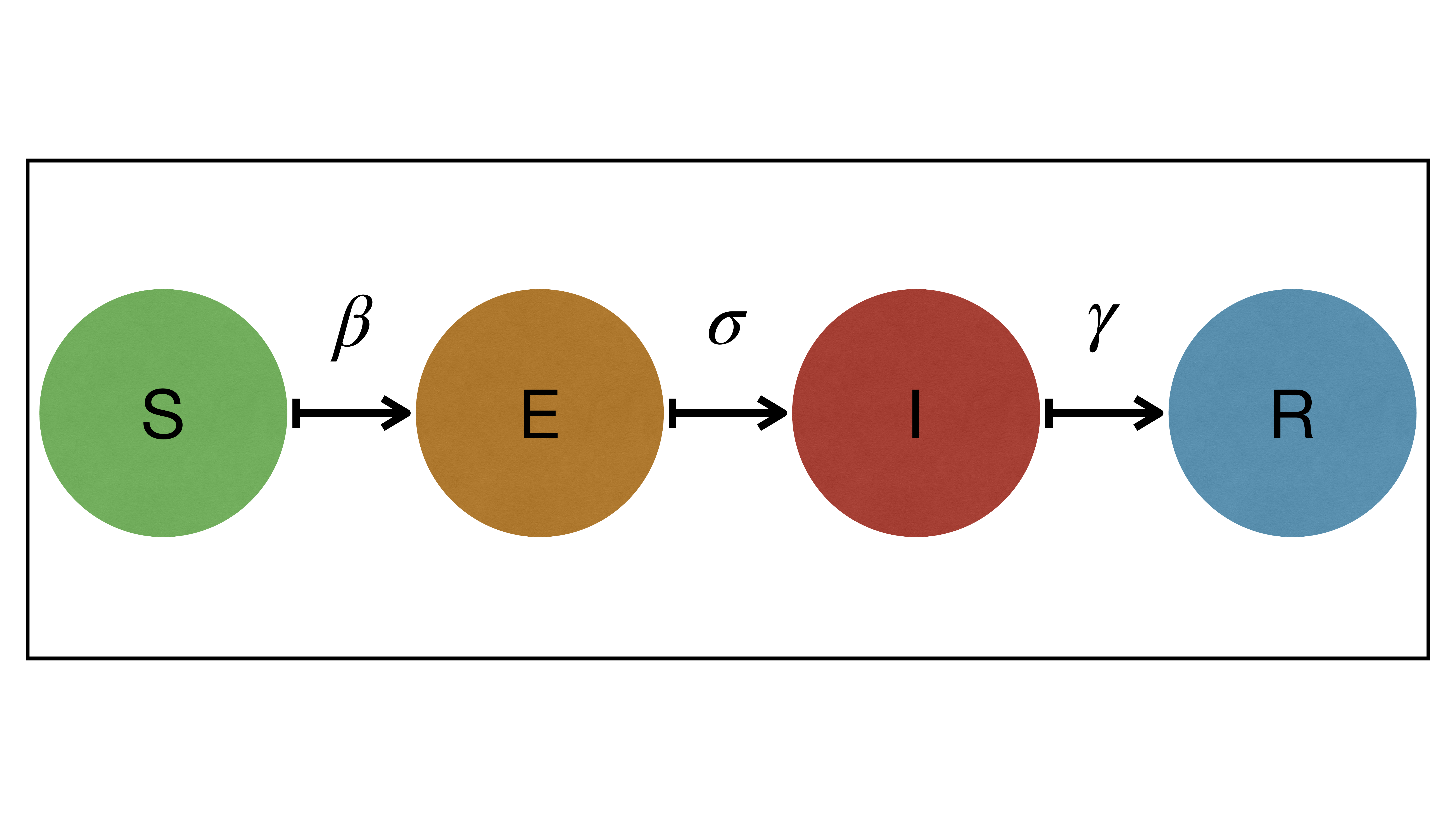}
\caption{\label{fig:SEIRa}}
\end{subfigure}
\begin{subfigure}{1\columnwidth}
\includegraphics[width = \textwidth]{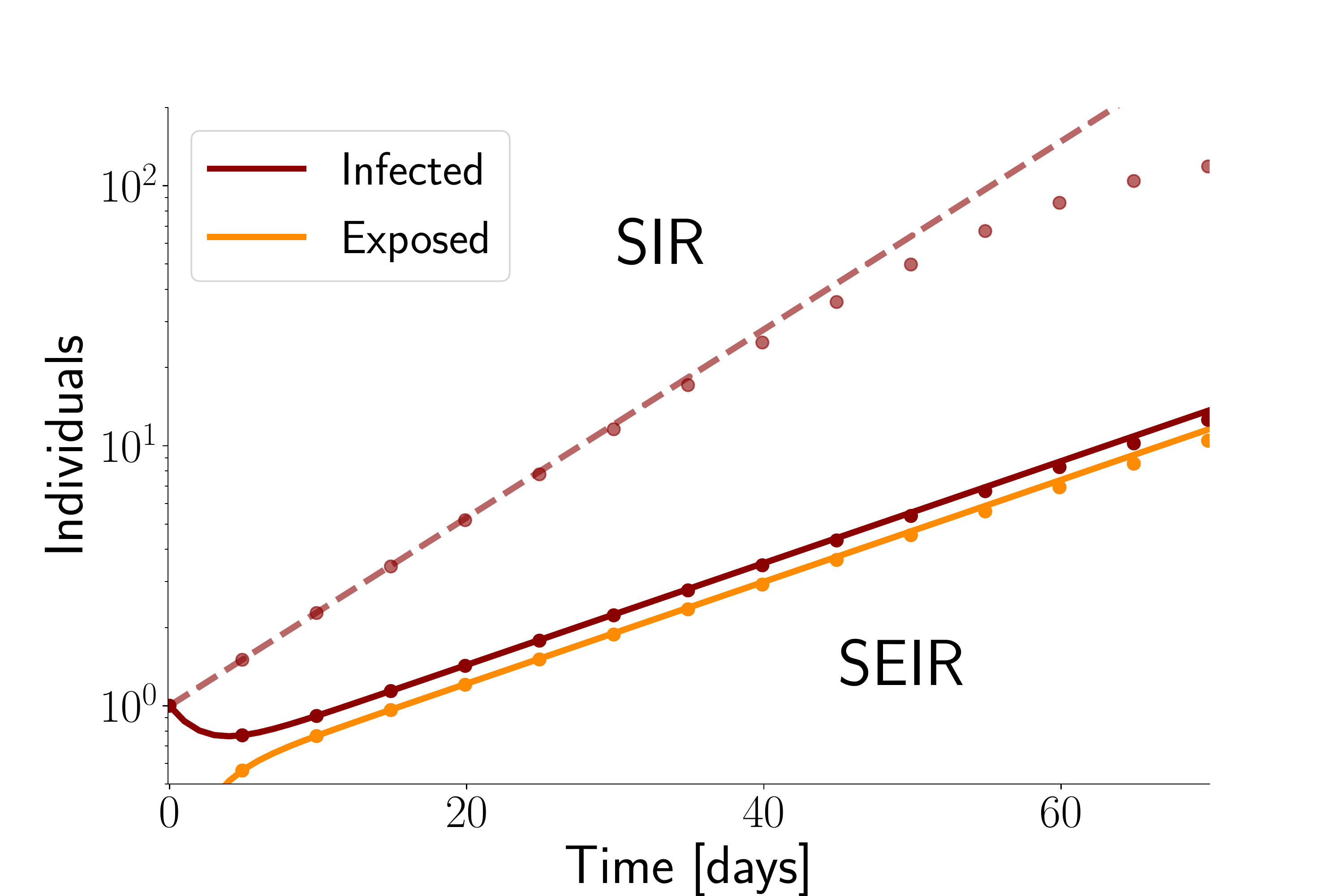}
\caption{\label{fig:SEIRb}}
\end{subfigure}
\caption{ \textbf{Deterministic dynamics of the SEIR model}. {\bf (a)} Schematic of the infection dynamics between 4 states: Susceptible, Exposed, Infected and Recovered individuals. The transition rates between these states, $\beta$, $\sigma$ and $\gamma$, enter the dynamical equations~(\ref{eq:ODE1})-(\ref{eq:ODE4}). {\bf (b)} Epidemic growth of exposed and infected populations. After an initial phase, both populations grow exponentially at the same rate $\lambda$, as given by eq.~(\ref{eq:lambda_SEIR}) (lines: analytical solution in the exponential growth regime, $S\approx N$; dots: numerical simulations). The corresponding SIR model at the same value of $R_0$ is shown for comparison (dashed lines); the simulation results show the onset of saturation by depletion of susceptible individuals. Parameters: $\beta = 0.25$, $\gamma = 1/6$, $\sigma = 1/4$ (for SEIR).}
\end{figure}

The most commonly used model in epidemiology is the classical SIR model~\cite{Bailey1975, keeling2008}. It has been used to study the behavior of epidemics because it captures most of the general features of the dynamics and requires few parameters. The SIR model consists of a population of individuals, each of which can be susceptible ($S$), infected ($I$) or recovered ($R$). In the following, we use an extension of the SIR model, the so-called SEIR model, which contains an additional compartment of exposed ($E$) individuals. The exposed compartment contains infected individuals which are not yet able to transmit the infection, because the pathogen is in an incubation or pre-contagious phase. The SEIR model has previously been used to model epidemics on contact networks in the context of the 2005 SARS-CoV outbreak~\cite{small2005} and to estimate the turning period of SARS-CoV-2 outbreaks~\cite{Yuan2021}.

\begin{figure*}[t!]
\begin{subfigure}{.49\textwidth}
\centering
\includegraphics[width = \textwidth]{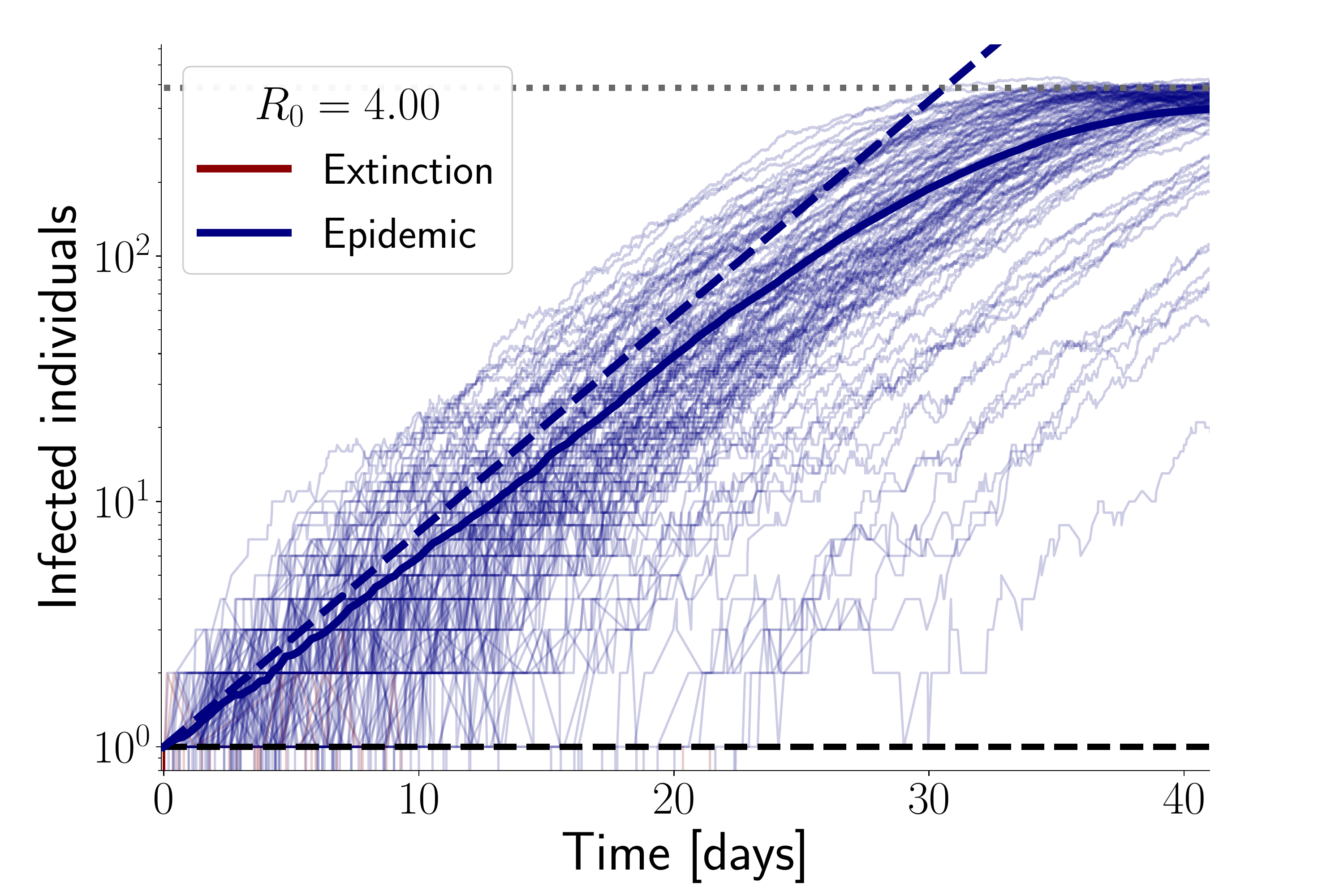}
\caption{\label{fig:stochasticity1}}
\end{subfigure}
\begin{subfigure}{.49\textwidth}
\centering
\includegraphics[width = \textwidth]{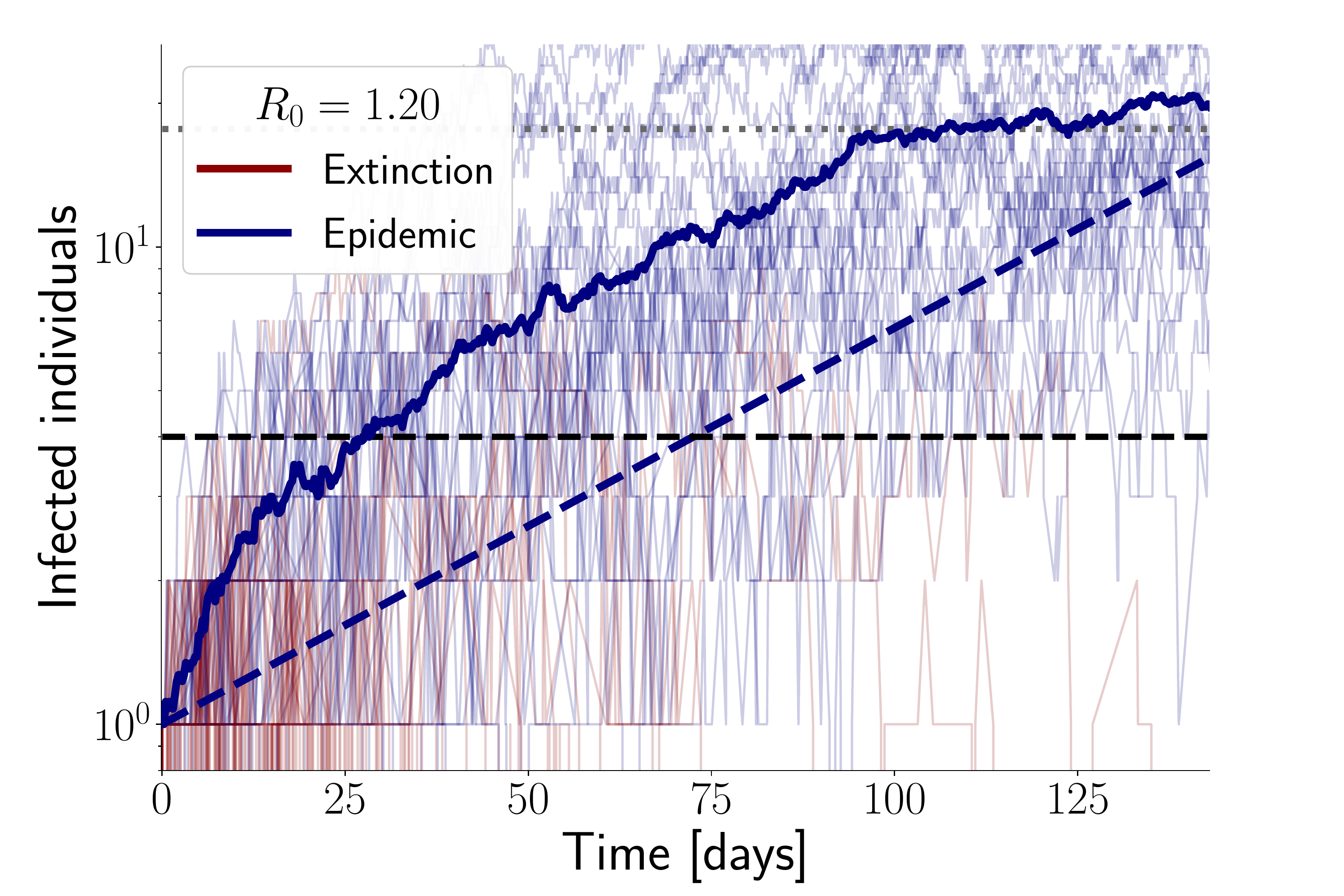}
\caption{\label{fig:stochasticity2}}
\end{subfigure}
\caption{\textbf{Stochastic dynamics of the SEIR model}. Time courses of the number of infected individuals are shown for outbreaks bound for extinction (red lines) and for proliferation (blue lines), together with the ensemble average of proliferation-bound outbreaks (thick blue lines) and the corresponding deterministic dynamics (dashed black lines). A threshold number of infected individuals, the so-called establishment size  $\nint^* (R_0)$ (dashed green lines), separates regimes predominantly extinction-bound and predominantly proliferation-bound time courses. The dynamics is shown for two reproductive numbers: {\bf (a)} $R_0 = 4.0$, {\bf (b)} $R_0 = 1.2$. With decreasing $R_0$, the establishment size increases and saturation effects ($S < N$; see Methods) cap the regime of exponential growth (dotted lines). Other simulation parameters: $N=5000$, $\sigma = 1/4$, $\gamma = 1/6$, {\bf (a)} $\beta = 0.66$, {\bf (b)} $\beta = 0.19$. 
\label{figs:stochasticity}}
\end{figure*}

The deterministic SEIR dynamics is governed by a set of differential equations,
\begin{eqnarray} 
\dot{S} &=&-\beta \frac{SI}{N}\label{eq:ODE1}\\
\dot{E} &=&\beta \frac{SI}{N} - \sigma E\label{eq:ODE2}\\
\dot{I} &=&\sigma E - \gamma I\label{eq:ODE3} \\
\dot{R} &=&\gamma I.
\label{eq:ODE4}
\end{eqnarray} 
These equations describe the infection dynamics in a well-mixed population of constant size $N$. Susceptible individuals get exposed at a rate proportional to the fraction of infected individuals with a transmission rate parameter $\beta$. Exposed individuals become infectious at a constant rate $\sigma$; this delay represents intra-host growth of the viral population up to a level where transmission to a next host becomes likely. Infected individuals recover at a constant rate $\gamma$ (Figure~\ref{fig:SEIRa}). In the model tuned to SARS-CoV-2, we use estimates of the average pre-contagious period ($\tau_\sigma\equiv \sigma^{-1} = $ 4 days) and infectious period ($\tau_\gamma\equiv \gamma^{-1} =$ 6 days), and we choose the parameter $\beta$ from the interval $0.075 -0.75$ days$^{-1}$~\cite{BarOn2020, He2020, Jones2021, Goyal2021}. A further, frequently used parameter to characterize epidemic growth is the so-called {\em reproductive number}, which is defined a the average number of transmissions from a single individual during its infectious period. In the well-mixed SEIR model, the reproductive number is simply related to the basic rate parameters, $R_0 = \beta / \gamma$~\cite{Allen2012}; however, this relation is no longer valid on a contact network. 

We are interested in the first stage of the dynamics, where the approximation $S \approx N$ is valid and the number of infected individuals grows exponentially with rate $\lambda$, as given by 
\begin{equation}
\lambda(R_0, \sigma, \gamma) =-\frac{\sigma + \gamma}{2} + \frac{1}{2}\sqrt{(\sigma - \gamma)^2 + 4\sigma\gamma R_0}
\label{eq:lambda_SEIR}
\end{equation}
(see Methods). Here we have used $R_0$ instead of $\beta$ as the independent parameter, because the relation (\ref{eq:lambda_SEIR}) remains valid on a contact network (see below). The reproductive number delineates the regimes of deterministic growth ($\lambda > 0$ for $R>1$) and decline ($\lambda < 0$  for $R_0<1$). 

In Figure~\ref{fig:SEIRb}, we show the deterministic  growth of an SEIR epidemic starting from a single infected individual. After a brief initial period, the infected and exposed populations grow at the same rate $\lambda$ and at comparable relative size. Because incubation slows down growth, this rate is lower than for a SIR model at the same value of $R_0 = \beta / \gamma$, which is given by $\lambda(R_0, \gamma)=(R_0 - 1)\gamma$ (dashed lines in Figure~\ref{fig:SEIRb}). In other words, in the deterministic dynamics, a sizeable exposed phase has a deleterious fitness effect for the pathogen. In the stochastic theory, as we discuss below, the exposed phase also generates a beneficial effect, because it reduces fluctuations leading to early extinctions.

\subsection*{Stochastic infection dynamics}

\begin{figure*}[t!]
\begin{subfigure}{.48\textwidth}
\centering
\includegraphics[width = \columnwidth]{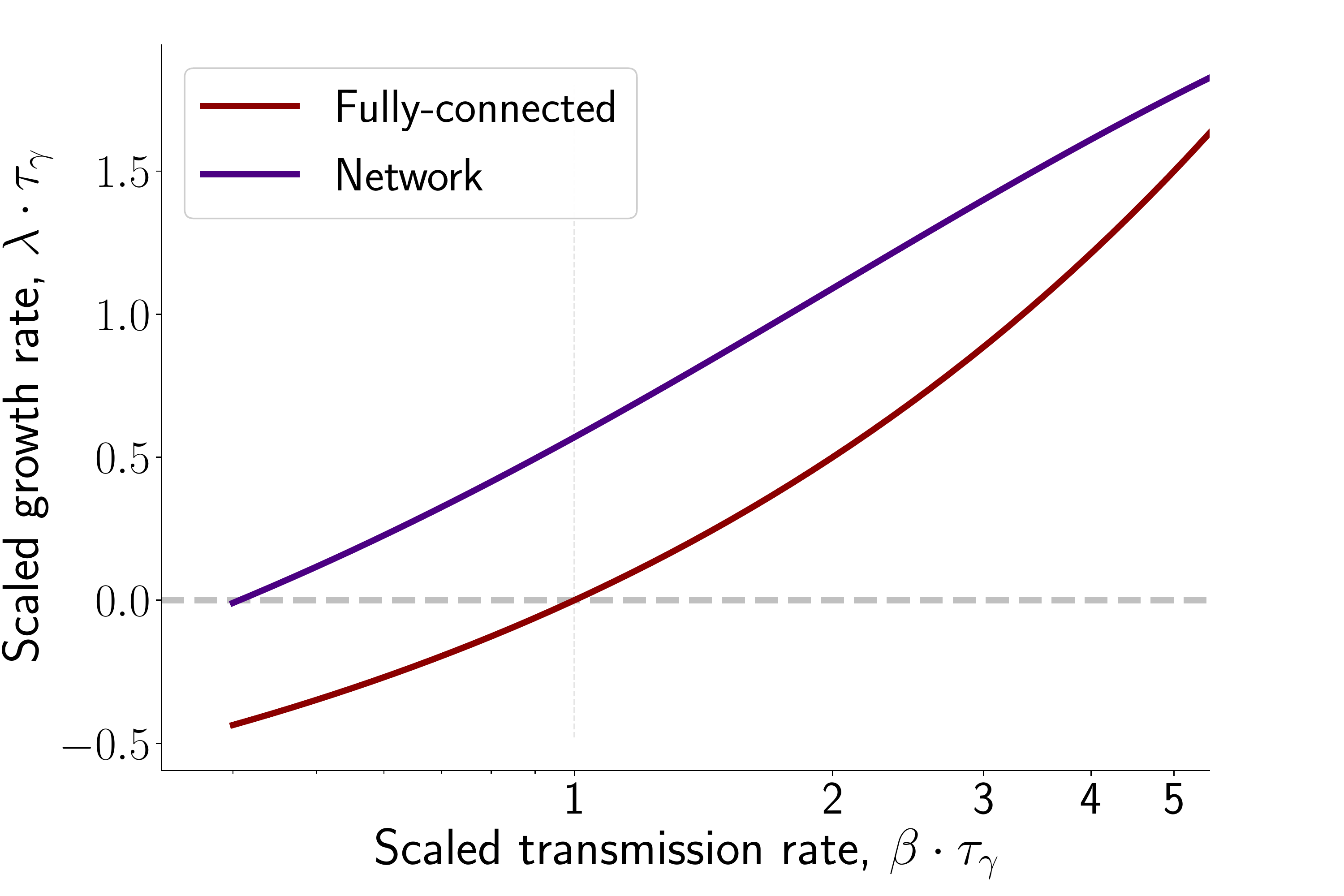}
\end{subfigure}
\begin{subfigure}{.48\textwidth}
\includegraphics[width = \columnwidth]{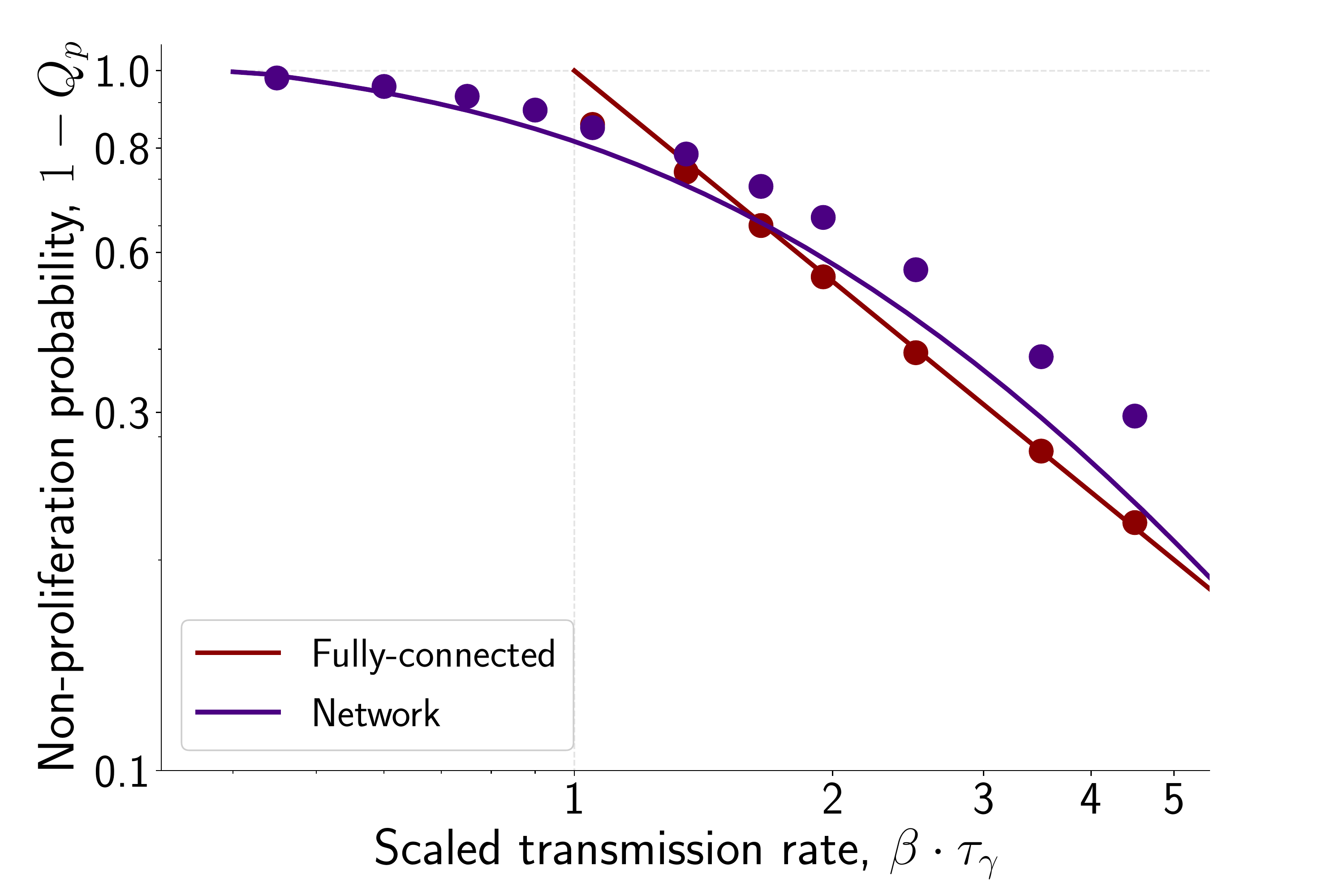}
\end{subfigure}
\caption{\textbf{SEIR epidemics on a contact network, compared to a well-mixed population.} {\bf (a)} Scaled growth rate, $\lambda /\gamma$, and {\bf (b)}  non-proliferation probability, $1 - Q_p$, on a network of contacts (purple) and in a homogeneous, fully connected system (red). Analytical results for the SEIR dynamics (lines) are compared to simulation results (symbols). Other simulation parameters: $N=2000$, $\alpha=2.68$ (network), $\sigma = 1/4$, $\gamma = 1/6$, $\beta = 0.075, 0.1, 0.12, 0.15$ (network), $0.17, 0.22, 0.27, 0.32, 0.42, 0.58, 0.75$ (network and well-mixed system). 
\label{fig:p_epi}}
\end{figure*}

Given initially low numbers of exposed and infected individuals, stochasticity plays an important role in the transmission dynamics of local outbreaks. In Figure~\ref{figs:stochasticity}, we plot time courses of the number of infected individuals from simulations of the SEIR model in a well-mixed population. Some of these time courses reach only small transient population sizes and then go extinct (red lines), the remainder proliferate to deterministic, initially exponential growth (blue lines). Three characteristics of these stochastic dynamics are readily recognized. First, the fraction, duration, and population size of extinction-bound time courses strongly depends on the reproductive number. Below, we calculate the proliferation probability $Q_p (R_0)$, given the initial condition of a single infected individual. Second, the fraction of extinction-bound trajectories decreases with increasing population size. Beyond a threshold $n^* (R_0)$, the so-called establishment size (black dashed lines in Figure~\ref{figs:stochasticity}), proliferation becomes the more likely fate. The establishment size will also be calculated from a suitably extended expression of the extinction probability. Third, in outbreaks bound for proliferation, the average number of infected individuals (thick blue lines) grows initially faster than expected from the deterministic dynamics (blue dashed lines). This initial boost is a posterior effect of sampling only time courses leading to epidemic growth; it is more pronounced for values of $R_0$ close to $1$ (Figure~\ref{fig:stochasticity2}). In this regime, typical stochastic trajectories reach the epidemic regime (black dashed lines) several days earlier than expected from the deterministic model. These stochastic initial-time effects can drastically affect the efficiency of monitoring and control strategies.

To study the stochastic dynamics of the SEIR model we assume that, at a given time, each individual $j$ in the populations is in a state $x_n\in\{s, e, i, r\} $. The state of an individual at time $t+dt$, given the state at time $t$, is governed by the transition probabilities
\begin{eqnarray} 
\mathcal{P}_j (s\rightarrow e) & =& \beta\frac{I}{N} dt 
\label{eq:StoE}
\\
\mathcal{P}_j (e\rightarrow i) & =& \sigma dt
\label{eq:EtoI}
 \\
\mathcal{P}_j (i\rightarrow r)  & =& \gamma dt.  
\label{eq:ItoR}
\end{eqnarray} 
In a well-mixed population, the transition rates are uniform across individuals and equal the deterministic rates in Eqs.~(\ref{eq:ODE1})-(\ref{eq:ODE4}). 

The proliferation probability of the well-mixed SEIR model can readily be computed from a branching process approximation with constant linear birth and death rates of infected individuals, which is appropriate in the exponential growth regime ($S\approx N$). The well-known result for the SIR model, $1 - Q_p = 1/R_0$ carries over to the SEIR model, because the probability of infections is independent of the incubation step~\cite{Allen2012, Allen2017}.

\begin{figure*}[t!]
\begin{subfigure}{.48\textwidth}
\centering
\includegraphics[width = 1\textwidth]{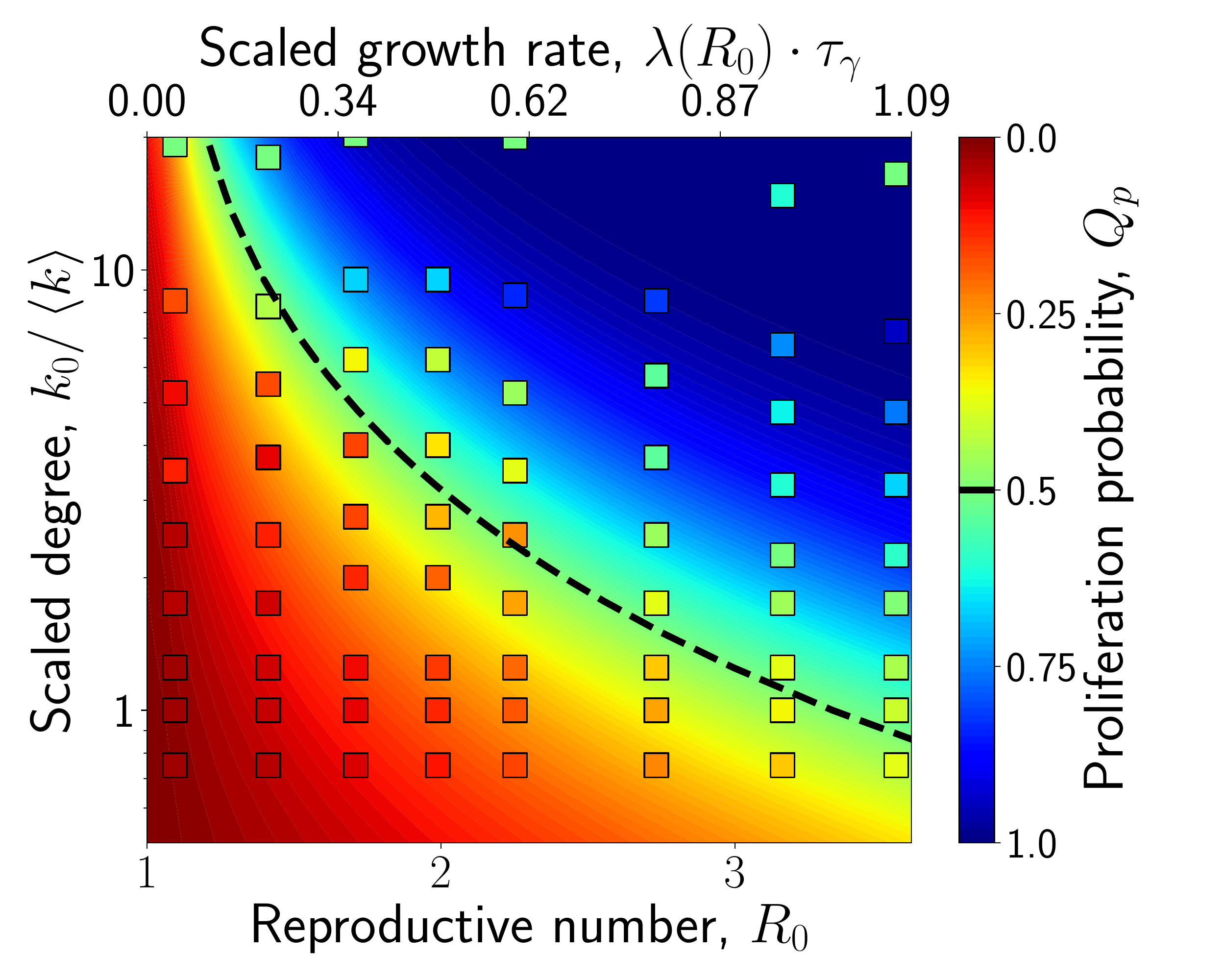}
\caption{\label{fig:p_epi_k0}}
\end{subfigure}
\begin{subfigure}{.48\textwidth}
\centering
\includegraphics[width = 1\textwidth]{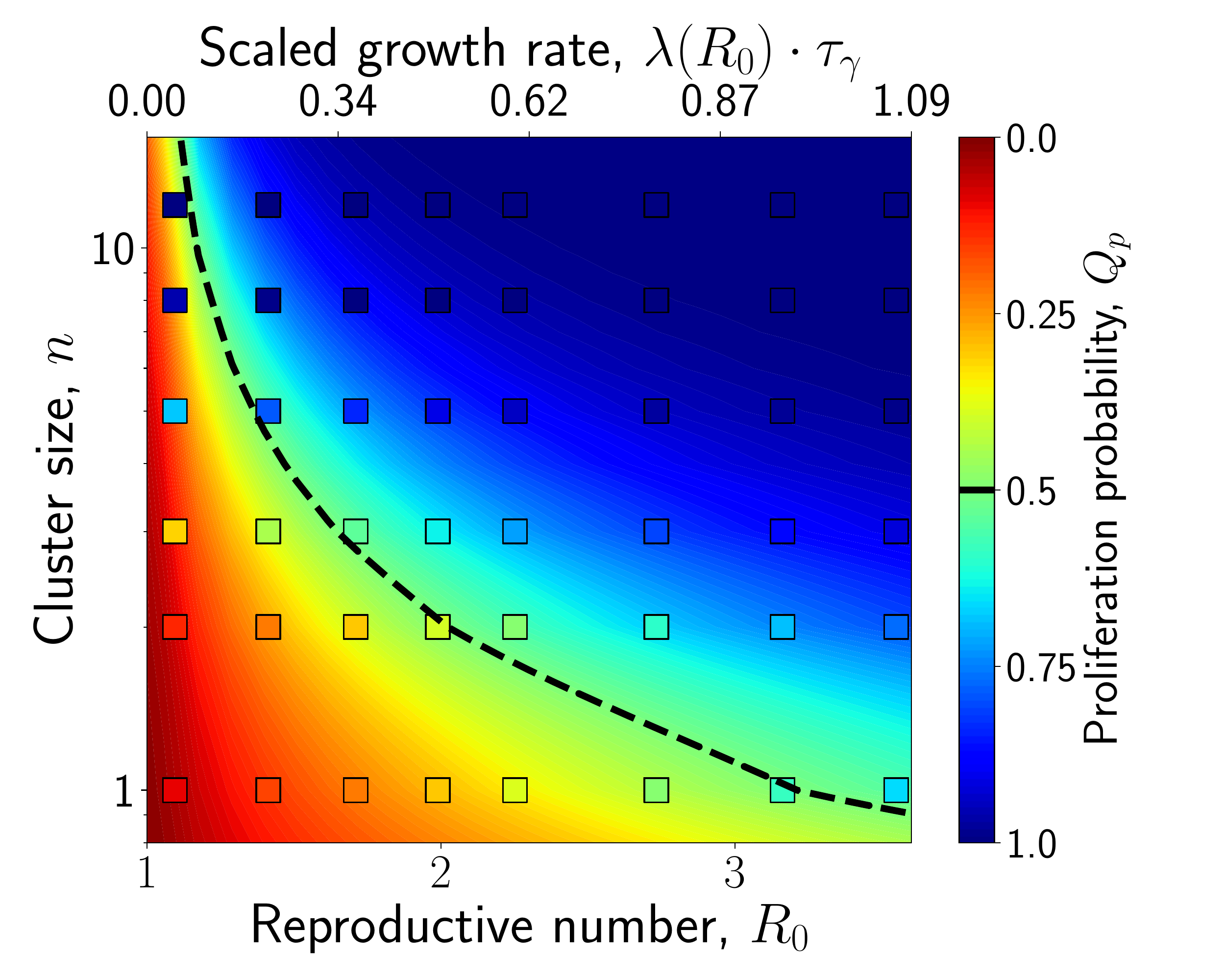}
\caption{\label{fig:p_epi_n}}
\end{subfigure}
\caption{\textbf{Stochastic outbreaks on contact networks.}  
{\bf (a)} Degree-dependent proliferation probability $Q_p (R_0, k_0)$ and threshold degree $k_0^* (R_0)$ (dashed line); simulation outcomes for various points in parameter space (squares) are compared to analytical predictions (background coloring). {\bf (b)} Conditional proliferation probability $Q_p (R_0, \nint)$  and establishment size $\nint^* (R_0)$ (dashed line). The top scale gives the scaled growth rate $\lambda/\gamma$ related to $R_0$ by eq.~(\ref{eq:lambda_SEIR}). Simulation parameters: $N=2000$, $\alpha=2.68$, $\sigma = 1/4$, $\gamma = 1/6$. 
 \label{figs:p_epi}}
\end{figure*}

To compute the establishment size, we evaluate the conditional proliferation probability $Q_p (R_0, \nint)$, given that the infected population has already reached a size $\nint$. This probability is a decreasing function of $\nint$, interpolating between the initial value $Q_p (R_0,\nint) = Q_p (R_0)$ and near-certain epidemic growth, $Q_p (R_0, \nint) \simeq 1$, of sufficiently large infected clusters (Figure~\ref{figs:stochasticity}). We obtain $Q_p (R_0, \nint) =  1 - \left(1/R_0\right)^{\nint}$, assuming that extinction from an initial state of $n$ infected individuals is equivalent to $n$ independent extinction processes, each of them starting from a single individual. The establishment size $\nint^* (R_0)$ is then obtained from the condition $Q_p = 1/2$, which gives $\nint^* (R_0) = \log{2}/\log{R_0}$ (black dashed lines in Figure~\ref{figs:stochasticity},~\ref{fig:p_epi_n} and~\ref{fig:sampling_cs}).

\subsection*{Infection dynamics on a contact network}

Multiple stochastic factors govern the initial stage of an outbreak~\cite{Newman2002, Lloyd-Smith2005, Allard2020, Althouse2020}. In addition to the intrinsic stochasticity of the dynamics given by Eqs~\ref{eq:StoE}--\ref{eq:ItoR}, there is heterogeneity in the initial condition of patient zero. Most studies have addressed differences in the reproductive number $R_0$ across the population, which can be  caused by a complex mixture of host, pathogen and environmental factors~\cite{Lloyd-Smith2005}. Here, we consider stochastic effects generated by a network of contacts that controls the interactions between individuals and translates into an effective variation of infectiousness (Figure~\ref{fig:network}). This (undirected) network is characterized by a broad degree distribution $\hat{p}(k)\sim k^{-\alpha}$ \cite{Barabasi1999}, which gives the probability that an individual in the network has $k$ contacts that can become transmission channels. While the degree distribution ignores the specific topology of the network, it captures the effect of largely connected nodes or hubs on the transmission statistics. Hubs play an important role in super-spreading events, which turns out to be the most relevant network feature for the degree-based testing protocol that we discuss below.

Given that infections take place between neighboring nodes of the contacts of the contact network, the SEIR transition probability from susceptible to exposed is given by 
\begin{equation}
\mathcal{P}_j (s\rightarrow i ) = \beta\frac{I_j}{ \langle k\rangle } dt,
\label{eq:StoE_N}
\end{equation}
where $I_j$ is the number of infected contacts (neighboring nodes) of node $j$ and $\left\langle\dots\right\rangle$ denotes averages over the degree distribution $\hat p(k)$. This local process replaces the infection step of Eq.~(\ref{eq:EtoI}) in a well-mixed population. The normalization factor $ \langle k\rangle$ in Eq.~(\ref{eq:StoE_N}) ensures that in the limit of large $ \langle k\rangle$ at constant $\beta$, the network dynamics converges to the well-mixed case. 

In Figure~\ref{fig:p_epi}, we compare the SEIR infection dynamics on a contact network with the dynamics in a well-mixed population at the same parameters $\beta$, $\gamma$, $\sigma$. In the parameter regime shown, the epidemic on a network grows faster; most strikingly, the regime of positive growth extends well below the growth threshold $\beta/\gamma =1$ in the well-mixed case (Figure~\ref{fig:p_epi}a). Consistently, the proliferation probability $Q_p$ remains positive in the same regime; for larger values of $\beta/\gamma$, however, proliferation is less likely on a network than in a well-mixed population. This pattern indicates two opposing effects of the contact network on epidemic growth. First, the epidemic spreads preferentially on high-connectivity nodes, which enhances the effective transmission rate. Second, infections deplete susceptible individuals specifically in the neighborhood of an infected node, leading to saturation effects at larger reproductive numbers.

\subsection*{Contact-dependent super-spreading} 

Most importantly, the proliferation probability of outbreaks on a contact network is no longer uniform but depends strongly on the contacts of the first infected individual (patient zero). In Figure~\ref{figs:p_epi}a, we plot $Q_p$ as a function of the degree of patient zero, $k_0$, and the reproductive number. Similar to the population threshold defined before, we define a threshold degree $k_0^* (R_0)$, such that $Q_p (R_0, k_0^*)=1/2$ (dashed line). As expected, highly connected seeding nodes in the network are more likely to produce epidemics than nodes with low connectivity. This is the so-called super-spreader effect: most of the epidemics are seeded by highly connected individuals. The effect is most pronounced for small values of $R_0$. Even for $ \beta/\gamma\lesssim1$, individuals with connectivity $\gtrsim10$-fold above average are likely to seed epidemics. These rare events correspond to infections seeded in a connectivity hub in the network. Conversely, individuals that have a connectivity smaller than the average connectivity in the population have an extinction probability close to 1 in the same regime. In the next section, we will use $Q_p(R_0, k_0)$ as a weighting factor for testing protocols. 

A similar pattern is observed in the conditional proliferation probability $Q_p (R_0, n)$, which depends on an intermediate outbreak size $n$ as defined above (Figure~\ref{fig:p_epi_k0}). This similarity is intuitively clear: an individual of degree $k_0$ generates, after one transmission step, an infected cluster of size $\nint$ proportional to $k_0$. 

In the remainder of this section, we sketch the derivation of the above results from the outbreak statistics on networks (more details are given in Methods; readers interested primarily in the application to contact protocols may skip this part). First, following ref.~\cite{Newman2002}, we define the basic building block of the epidemic dynamics on a network: the probability $T$ that an infected individual transmits the infection to a given contact; in other words, the probability that the infection spreads along a given branch of the network.  This probability determines the reproductive rate on the network, 
\begin{eqnarray} 
R_0(T) &=& \langle \! \langle (k-1) \, T \, \rangle \! \rangle =  \frac{\left\langle k(k-1)\right\rangle}{\left\langle k\right\rangle} \, T.
\label{eq:R0_N}
\end{eqnarray}
Here $\langle \dots \rangle$ denotes averaging over the degree distribution $\hat p(k)$ and $\langle \! \langle \dots \rangle \! \rangle$ averaging over the nearest-neighbor degree distribution $\hat p_2(k)=k\hat p(k)/\left\langle k\right\rangle$. The resulting growth rate in the exponential regime, $\lambda (R_0, \gamma, \sigma)$ is then given by eq.~(\ref{eq:lambda_SEIR}). 

The branch transmission probability $T$, which has been used as an independent parameter in previous work, has no direct analogue in a well-mixed system. However, we can express $T$ in terms of the rate parameters $\beta$ and $\gamma$, 
\begin{equation}
 T(\beta, \gamma)=\frac{1}{1+\frac{\langle k\rangle}{\beta/\gamma}}, 
 \label{eq:T}
\end{equation}
where we have used Eq.~(\ref{eq:StoE_N}) (see Methods). Together with Eq.~(\ref{eq:R0_N}), we obtain 
\begin{equation}
R_0(\beta, \gamma)=\left(\frac{\left\langle k^2 \right\rangle - \left\langle k\right\rangle}{\langle k\rangle ^2 + \frac{\beta}{\gamma}\left\langle k\right\rangle}\right) \frac{\beta}{\gamma}; \nonumber
 \label{eq:R0_N_2}
\end{equation}
this relation compares the reproductive number on a contact network with the corresponding well-mixed system, which has $R_0 = \beta/\gamma$. It displays the two network effects discussed above: preferential spreading on high-degree nodes and saturation of susceptible contacts (Figure~\ref{fig:p_epi}b). In the well-mixed limit, $(\langle k \rangle \gg1$ with $\mathrm{Var}(k) = o(\langle k \rangle^2))$, we get $R_0 \rightarrow \beta/\gamma$ as expected. 

\begin{figure}[h!]
\centering
\begin{subfigure}{.98\columnwidth}
\includegraphics[width = \textwidth]{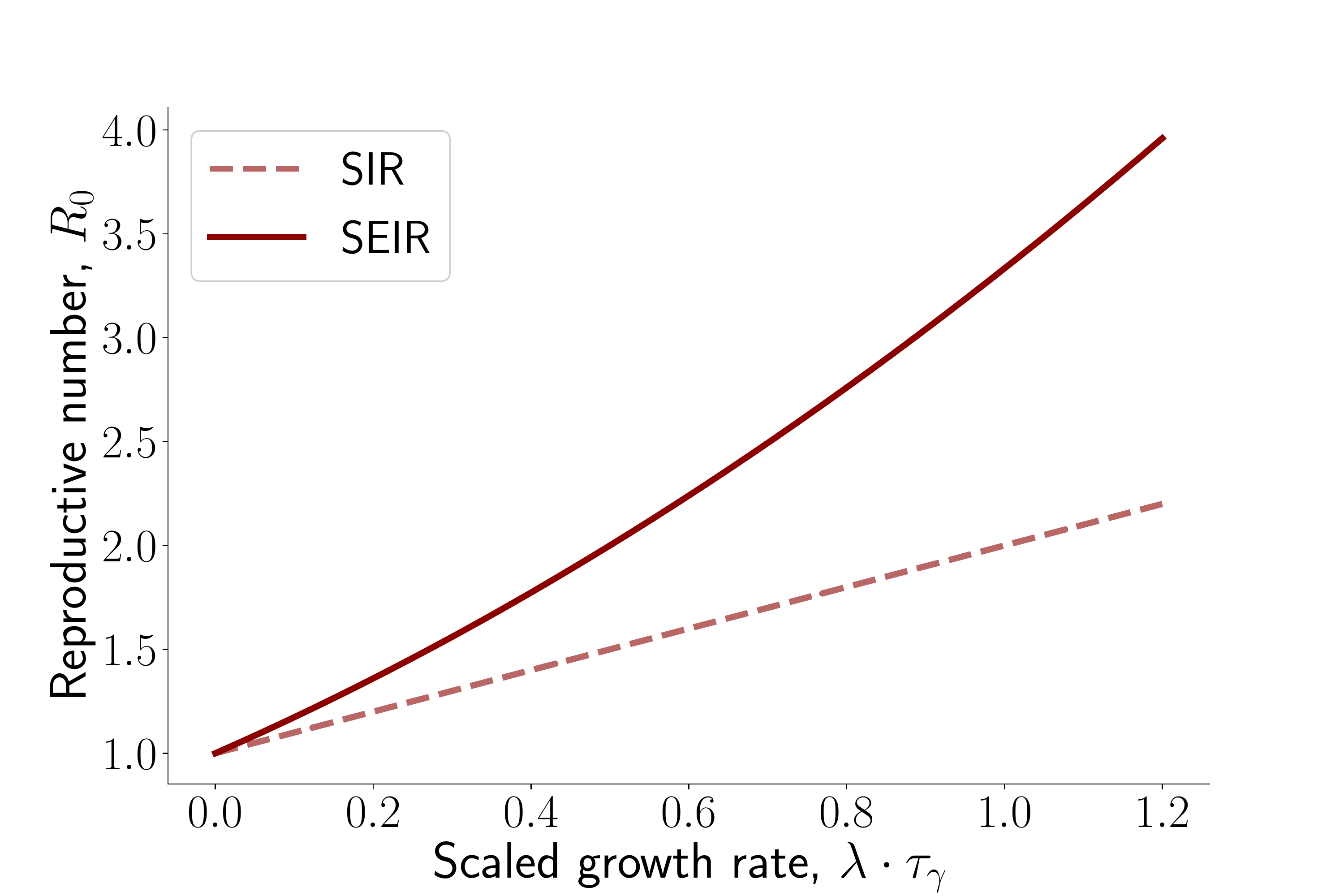}
\caption{}
\label{fig:inference1}
\end{subfigure}
\begin{subfigure}{.98\columnwidth}
\includegraphics[width = \textwidth]{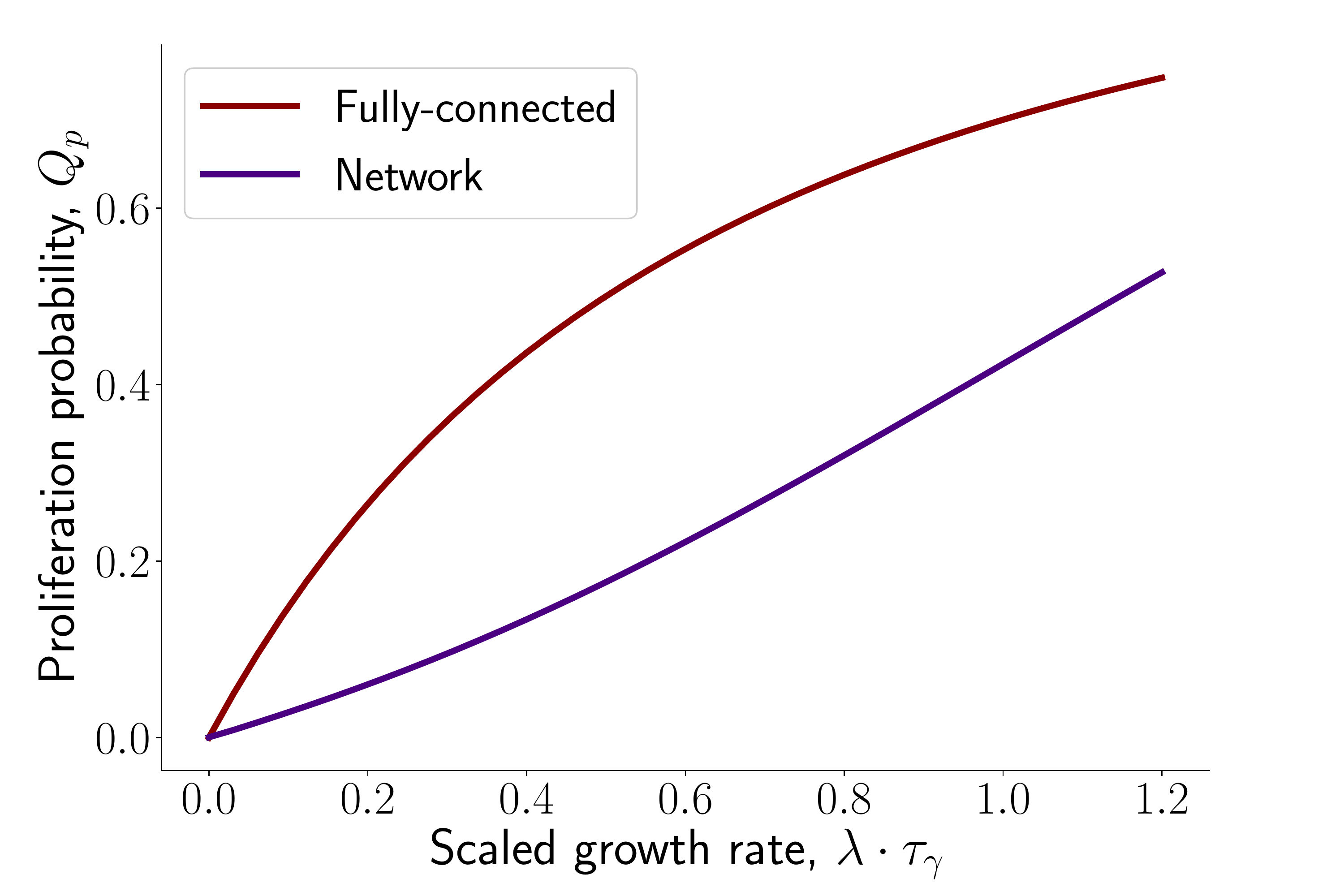}
\caption{}
\label{fig:inference2}
\end{subfigure}
\caption{{\bf Inferring reproductive number and proliferation probability.} Inference procedures can be based on three observed parameters: growth rate, $\lambda$, and average incubation and infectivity periods, $\tau_\gamma $ and $\tau_\sigma$. {\bf (a)} At given parameters, the SEIR model has a higher $R_0$ than the SIR model. {\bf (b)} At given parameters, the network model has a lower proliferation probability than the well-mixed model. }
\label{figs:inference}
\end{figure}

\begin{figure*}[t!]
\centering
\begin{subfigure}{.49\textwidth}
\centering
\includegraphics[width=\textwidth]{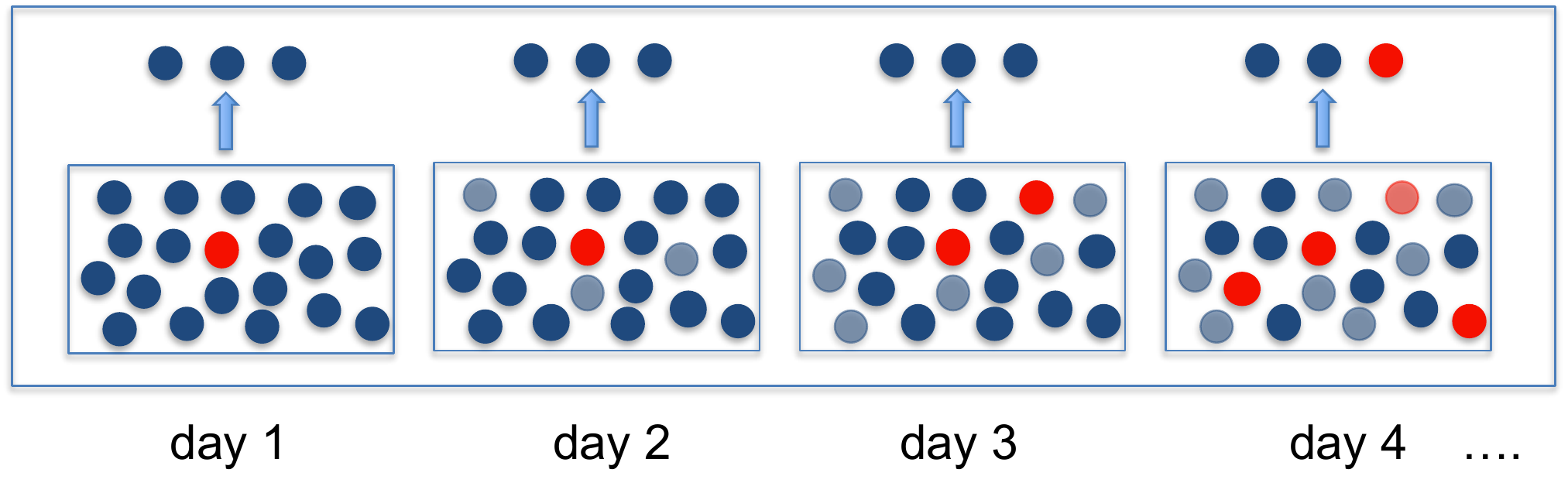}
\caption{}
\label{fig:sampling}
\end{subfigure}
\begin{subfigure}{.49\textwidth}
\centering
\includegraphics[width = \textwidth]{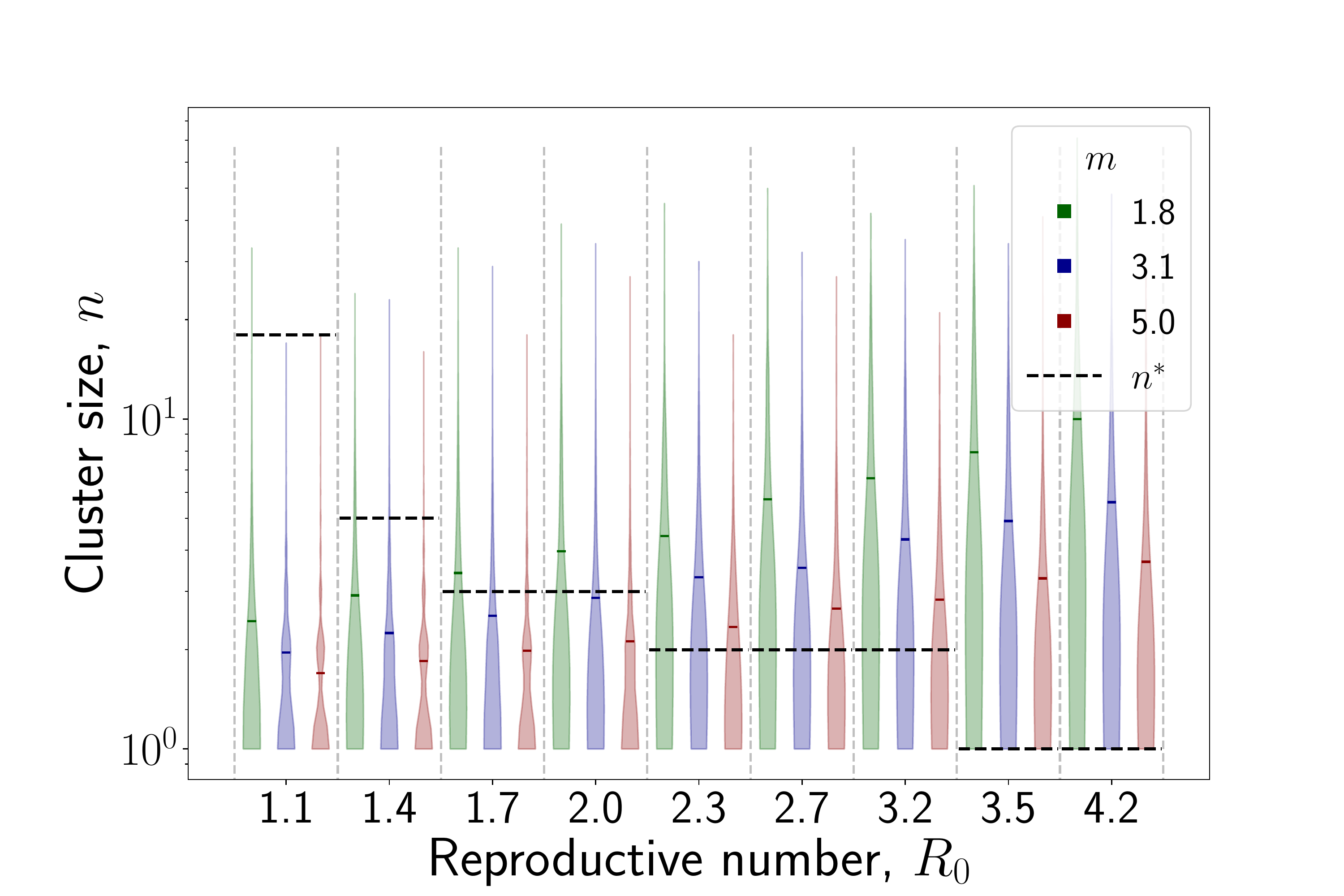}
\caption{}
\label{fig:sampling_cs}
\end{subfigure}
\caption{{\bf Surveillance by random testing.} {\bf (a)}  Daily random testing protocol. A constant number of individuals are chosen randomly everyday for testing. In addition, individuals that are selected, are not replaced in the sampling pool for the next $\tau_{\sigma}$ days (light blue dots).  The outbreak starts with one infected individual (red dots) at day one. {\bf (b)} Performance of the random testing protocol in terms of the infectious cluster size at the time of a first detection. Black horizontal lines show the establishment threshold $\nint^*$.\label{figs:surveillance}} 
\end{figure*}

Next, we use the branch transmission probability $T$ to compute the probability that an outbreak proliferates, given that it originates on a node of degree $k_0$. Again, a previous result for the SIR model ~\cite{Newman2002, Meyers2005} carries over to the SEIR model, because incubation delays growth but does not affect the probability of transmission. The proliferation probability takes the form 
\begin{equation}
Q_p (R_0, k_0) = 1- \big [1-T(R_0)+ q_b T(R_0) \big ]^{k_0}. 
\label{eq:p_ext_k0}
\end{equation}

Here $q_b$ is the probability that a sub-outbreak originating on a given branch of the network does not proliferate; this probability can be obtained from a self-consistent summation procedure (ref.~\cite{Newman2002}, see Methods). The product on the r.h.s.~of eq.~(\ref{eq:p_ext_k0}) says that transmissions along each of the $k_0$ branches originating from patient zero are statistically independent sub-outbreaks. The relation between $T$ and $R_0$ is given by Eq.~(\ref{eq:R0_N}). 

The global proliferation probabilities are then given by averages over the contact network~\cite{Newman2002, Meyers2005}. We obtain 
\begin{equation} 
Q_p(R_0)  =  
 \left\langle Q_p (R_0, k)\right\rangle
 \label{eq:p_ext_N}
\end{equation}
and 
\begin{equation}
Q_p (R_0, \nint) = 1-\left\langle\!\left\langle 1- Q_p (R_0, k-1) \right\rangle\!\right\rangle^{\nint}. 
\label{eq:p_ext_n}
\end{equation}
The product on the r.h.s.~of eq.~(\ref{eq:p_ext_n}) says that transmissions originating from each of the $n$ initially infected individuals are statistically independent sub-outbreaks; this form is analogous to eq.~(\ref{eq:p_ext_k0}). A given individual has been infected by one of its contacts and has $(k-1)$ contacts left to transmit the infection further. We note that these expressions take into account the contacts of typical individuals in the outbreak rather than individual contacts of specific individuals; for this reason, the average is taken using the nearest-neighbor degree distribution $\hat p_2(k)$~\cite{Newman2002}. As before, the conditional proliferation probability determines the establishment size of epidemics in a contact network, $\nint^*(R_0)$, by the condition  $Q_p(R_0, \nint^*) =1/2$. Below, we use this function to characterize the performance of surveillance testing protocols.

\subsection*{Inference of outbreak statistics from empirical data} 

The structures of the epidemiological model and of the contact network have important implications for the analysis of empirical data, in particular, for predicting the likelihood of future local outbreaks. 

First, the reproductive number $R_0$ is often inferred from the observed growth rate $\lambda$. In this procedure, we have to take into account infection and incubation periods, as given by the SEIR relation in Eq.~(\ref{eq:lambda_SEIR}). Using the corresponding SIR relation, which neglects incubation, the reproductive number can be drastically underestimated (Figure~\ref{fig:inference1}). 

Second, given a correctly inferred value of $R_0$, the contact network shapes the corresponding proliferation probability of local outbreaks, as given by eqs.~(\ref{eq:p_ext_k0}) - (\ref{eq:p_ext_n}). Using the naive expression for a well-mixed system, $Q_p = 1/R_0$, proliferation can be substantially overestimated (Figure~\ref{fig:inference2}); the same applies to the conditional probabilities $Q_p (R_0, k_0)$ and $Q_p (R_0, n)$. The correct inference of $R_0$ and of proliferation probabilities is a crucial step in the surveillance and containment protocols discussed below. 

\section*{Surveillance of local outbreaks}

\begin{figure*}[t!]
\centering
\begin{subfigure}{.48\textwidth}
\includegraphics[width = 1\textwidth]{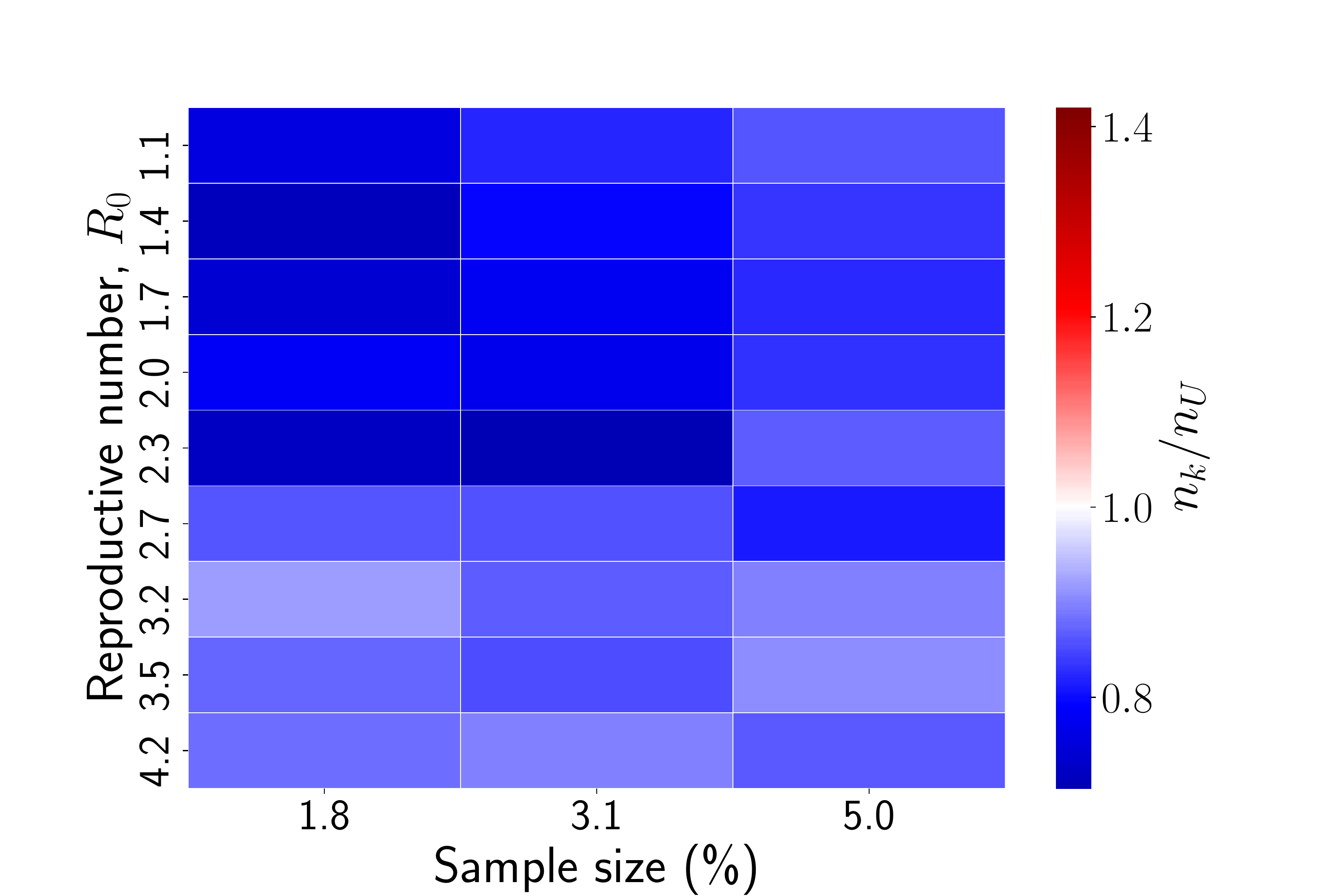}
\caption{}
\label{fig:sampling_cs_N}
\end{subfigure}
\begin{subfigure}{.48\textwidth}
\includegraphics[width = 1\textwidth]{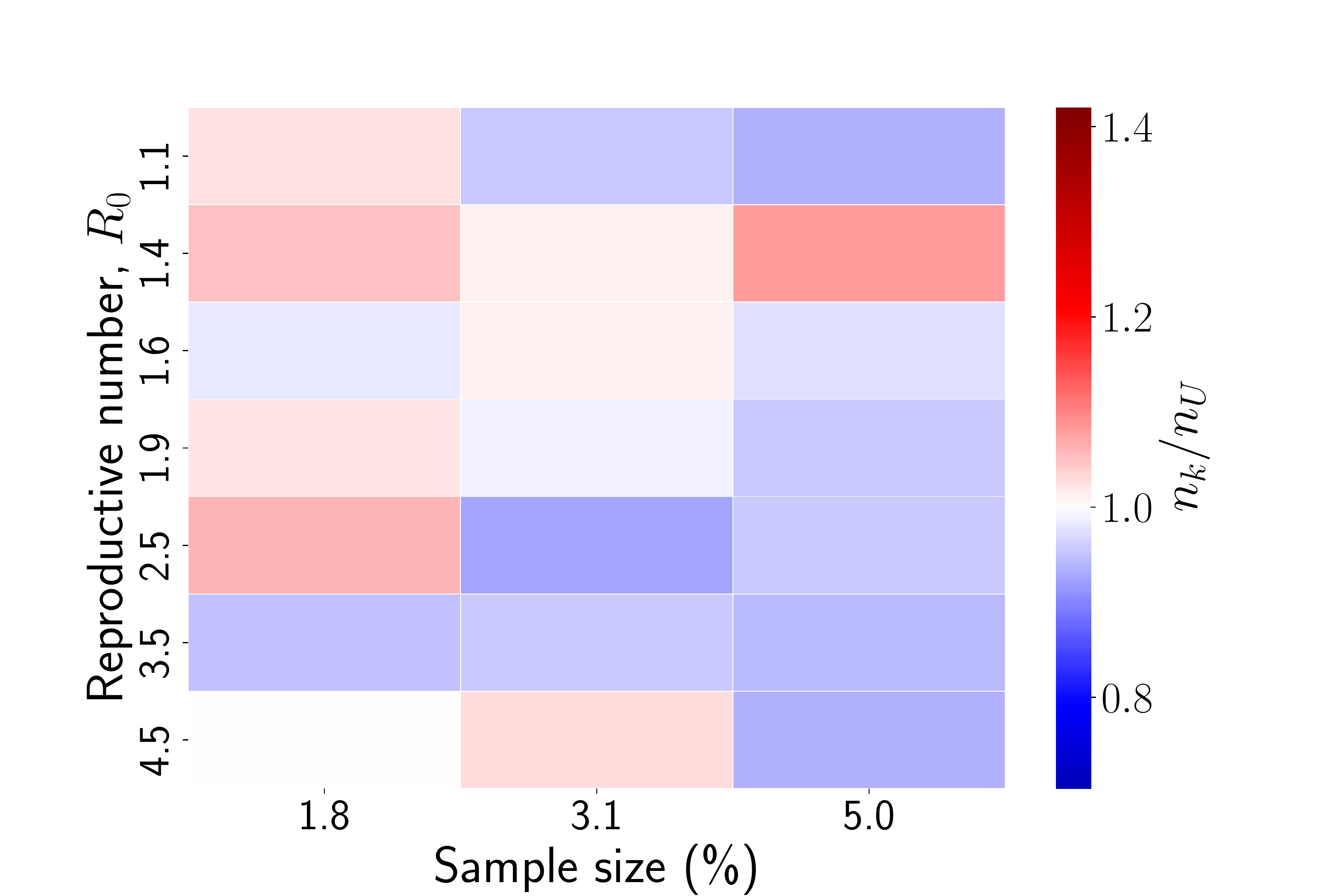}
\caption{}
\label{fig:sampling_cs_FC}
\end{subfigure}
\caption{{\bf Degree-based random testing protocol.} Performance of the degree-based random testing protocol relative to the uniform protocol in terms of infected cluster size $n$ at the time of detection; we show the case of individuals interacting {\bf (a)} under a network of contacts and {\bf (b)} in a well-mixed population as a control.\label{figs:degree_based}}
\end{figure*}

At early stages of an outbreak, the number of infected individuals can grow substantially before a first individual, if any, develops symptoms. For this reason, a surveillance strategy based only on symptomaticity will be suboptimal. Here we propose a surveillance strategy based on a daily random testing protocol of a fraction of the total asymptomatic population to detect infected individuals and prevent epidemics. 

We start by considering a simple protocol that consists of a daily uniform sampling of $m$ random asymptomatic individuals from the total population. Sampled individuals are removed from the testing pool for the next $\tau_\sigma$ days (Figure~\ref{fig:sampling}). We choose the time period of no replacement because if an individual is tested negative and gets infected on the next day, that individual will not test positive for the next $\tau_\sigma$ days on average. After this period of time, tested individuals are included in the testing pool again.

To evaluate the performance of the monitoring protocol, we focus on the average cluster size of infected individuals at the moment of detection and compared with the cluster size establishment limit defined above.  Figure~\ref{fig:sampling_cs} shows that the initial reproductive number is a crucial parameter if we want to detect outbreaks before they are likely to proliferate. For an initial reproductive number close to 1, even for the smallest sampling size considered here ($1.8\%$), the average outbreak size at the first detection is smaller than the establishment size. However, for values of $R_0\approx 2.0$, a sampling size larger than $5\%$ is required to achieve containment. For larger values of $R_0$, where the establishment size is approximately 1 individual (for $R_0>2.0$), other criteria need to be considered to quantify the performance of the surveillance protocol. For example, the time of first detection is crucial to minimize the epidemic size.

\subsection*{Contact-dependent monitoring protocol}

As mentioned above, the degree of the seeding node has a big impact on the probability of having an epidemic. Using Bayes' rule, we calculate the probability $\hat{p}(k_0|\mathrm{proliferation})$ that an epidemic originates from a node of degree $k_0$, 
\begin{equation}
\hat{p}(k_0|\mathrm{proliferation}) \sim \frac{Q_p (R_0, k_0)}{Q_p (R_0)}.
\label{eq:loglike1}
\end{equation}

We upgraded the testing surveillance by using the weight function $\omega(k_0) \equiv \hat{p}(k_0|\mathrm{proliferation}) $ for the random sampling of the protocol. With this weighting, we test highly connected individuals more often than poorly connected ones (Figure~\ref{fig:loglikelihood} in Methods).

We find an appreciable improvement of the degree-based testing protocol compared to the uniform testing protocol. The performance is better both in terms of the cluster size of the outbreak (Figure~\ref{figs:degree_based}) and in the time of detection (see Figure~\ref{figs:degree_based2} in Methods). Consistently, the improvement is smaller for larger values of $R_0$, where the effects of the network are less significant.

Figure~\ref{figs:degree_based} shows the ratio between the cluster size at the moment when the first infected individual is detected in the degree-based sampling protocol, $n_k$, and in the uniform sampling protocol, $n_U$. Figure~\ref{fig:sampling_cs_FC} shows that when the population is well-mixed, there is no significant difference between the two protocols. However, when the network of contacts regulates the interactions between individuals, as shown in Figure~\ref{fig:sampling_cs_N}, the expected cluster size is reduced in all the range of parameters that we explore. This reduction is greater for smaller values of $R_0$, reaching values larger than 20\%. We also measure a greater improvement for smaller sample sizes in the protocol, consistent with the fact that with larger uniform random samples it is more likely to detect infectious hubs that drive the proliferation. 

\section*{Discussion} 

To prevent large epidemics of infectious diseases, it is important to properly understand the dynamics at early stages of infectious outbreaks. When the average pre-infectious and incubation periods are similar to the average infectious period, it is important to consider a SEIR model rather than an SIR model. 
The early stage of the outbreak, when $S\approx N$, requires an appropriate stochastic description of the dynamics, where extinction and establishment events arise as novel features compared to a deterministic description. The fact that an outbreak can go extinct can be used to set up NPIs under limited resources or logistic capacities, like the testing protocol presented here. Specifically, as we focus our attention in the regime of few exposed or infected individuals emerging from a single new infection within a small population, the establishment size of an outbreak, as defined above, is a good estimate for the maximum number of infected individuals that can be tolerated before detection of the outbreak. For different regimes of an epidemic, some of our approximations are no longer valid. For instance, we do not consider the case when the number of infections grows to the same order as the size of the population. In this case, the proliferation probability becomes large and the surveillance testing protocol is no longer a suitable strategy for containment.

We have shown that the social connectivity structure is important for efficient monitoring of infectious outbreaks. The variability of the effective reproductive number introduced by a contact network in the population considerably impacts on the outbreak dynamics. More specifically, a heavy-tailed degree distribution creates a substantial super-spreader effect, by which highly connected individuals are more likely to start outbreaks that reach establishment than poorly connected individuals. The super-spreader effect has been extensively studied in the literature~\cite{Goyal2021, Lloyd-Smith2005, Althouse2020, Reich2020, Miller2020}. Here we establish quantitative relations between degree distribution of the contact network and proliferation probabilities of outbreaks that are relevant for pre-emptive monitoring. Specifically, we obtain a connectivity regimes for patient zero where an outbreak is likely to proliferate or not.

Based on the contact structure of the population, we show that a degree-based random testing protocol outperforms an uninformed uniform testing protocol (but we do not claim this protocol is optimal in an absolute sense). The difference in performance depends on the specific parameters of the pandemic spread, which also has important consequences for the global control of pandemics. In practice, the implementation of a degree-based random testing protocol requires a proper estimation of the connectivity of individuals, which can be done by low-cost methods in terms of resources and logistics~\cite{Killworth1990, Russell1991, McCormick2010, Maltiel2015}. Moreover, the posterior collection of combined data of contact networks and the reconstruction of infectious outbreaks in the corresponding structured population~\cite{Lokhov2014, Altarelli2014, Lokhov2015} could be used to validate our sampling protocol. Together, this could help to better understand the sources of stochasticity in local outbreaks and to develop better specific NPIs.

\section*{Acknowledgments}
This work has been supported by the Deutsche Forschungsgemeinschaft (DFG) through the Collaborative Research Center 1310, Predictability in Evolution. We thank M. Meijers and D. Trimcev for a careful reading of the manuscript and R. S. McGee for making {\it Seirsplus} available.

\appendix
\section*{Methods  \label{appendix}}

\subsection*{Deterministic growth}

We use a classical epidemiological model consisting of susceptible $(S)$, exposed $(E)$, infected $(I)$ and recovered $(R)$ individuals.  Exposed individuals are infected individuals that do not test positive and cannot infect other individuals yet. The total population size is $N$. Here we consider the deterministic dynamics of the system described by the ODE system of Eqs.~\ref{eq:ODE1}-~\ref{eq:ODE4}. We are interested in the first stage of the dynamics, where the approximation $S \approx N$ is valid. With this approximation, and assuming the initial conditions $S(t)= N-1$, $E(t)=R(t)=0$ and $I(t)=1$, Eqs.~\ref{eq:ODE2} and \ref{eq:ODE3} can be written as
\begin{eqnarray} 
\dot{E}(t) &= -\sigma E + \beta I\\
\dot{I}(t) &= \sigma E - \gamma I
\label{eqs:ODE_SN}
\end{eqnarray} 
The solution of the system of Eqs.~\ref{eqs:ODE_SN} is given by 
\begin{equation}
\begin{pmatrix}
E(t) \\
I(t)
\end{pmatrix} = C e^{\lambda_+t}
\begin{pmatrix}
1\\
\frac{\lambda_++\sigma}{\beta}
\end{pmatrix}
- Ce^{\lambda_-t}
\begin{pmatrix}
1\\
\frac{\lambda_-+\sigma}{\beta}
\end{pmatrix}
\label{eq:E_I_sol}
\end{equation}
where the constant $C$ is determined by the initial condition and $\lambda_+$ and $\lambda_-$ are the eigenvalues of the matrix
\begin{equation}
\begin{pmatrix}
-\sigma & \beta \\
\sigma & -\gamma
\end{pmatrix}.
\end{equation}
that are given by
\begin{equation}
\lambda_\pm =-\frac{\sigma + \gamma}{2} \pm \frac{1}{2}\sqrt{(\sigma - \gamma)^2 + 4\sigma\beta}.
\end{equation}
As soon as the outbreak is seeded, the time evolution of $E$ and $I$ is lead by the first term in Eq.~\ref{eq:E_I_sol} because $\lambda_-<0$. With the initial conditions $E(0)=0$ and  $I(0)=1$, Eq.~\ref{eq:E_I_sol} can be written as
\begin{equation}
\begin{pmatrix}
E(t) \\
I(t)
\end{pmatrix} 
= \frac{\beta}{\sqrt{(\sigma - \gamma)^2 + 4\sigma\beta}} 
\begin{pmatrix}
1\\
\frac{\lambda_++\sigma}{\beta}
\end{pmatrix}
e^{\lambda_+ t}
\label{eq:E_I_sol2}
\end{equation}
As mentioned above, we decided to include the exposed compartment of the population because it affects dramatically the dynamics  of the infected individuals (Figure~\ref{fig:SEIRb}). More concretely, for SARS-Cov-2 the estimated average time that takes an exposed individual to become infected $\tau_\sigma$ is in the same order of magnitude of other relevant time-scale, namely the average infectious period $\tau_\gamma$~\cite{BarOn2020}. Additionally, under these circumstances the classical relation between the exponential growth rate and the parameter $R_0$, the initial reproductive number, defined as the quotient between the infectious rate and the recovery rate $\beta/\gamma$, need to be revisited. 

\begin{figure*}[th!]
\begin{subfigure}{.44\textwidth}
\centering
\includegraphics[width = \textwidth]{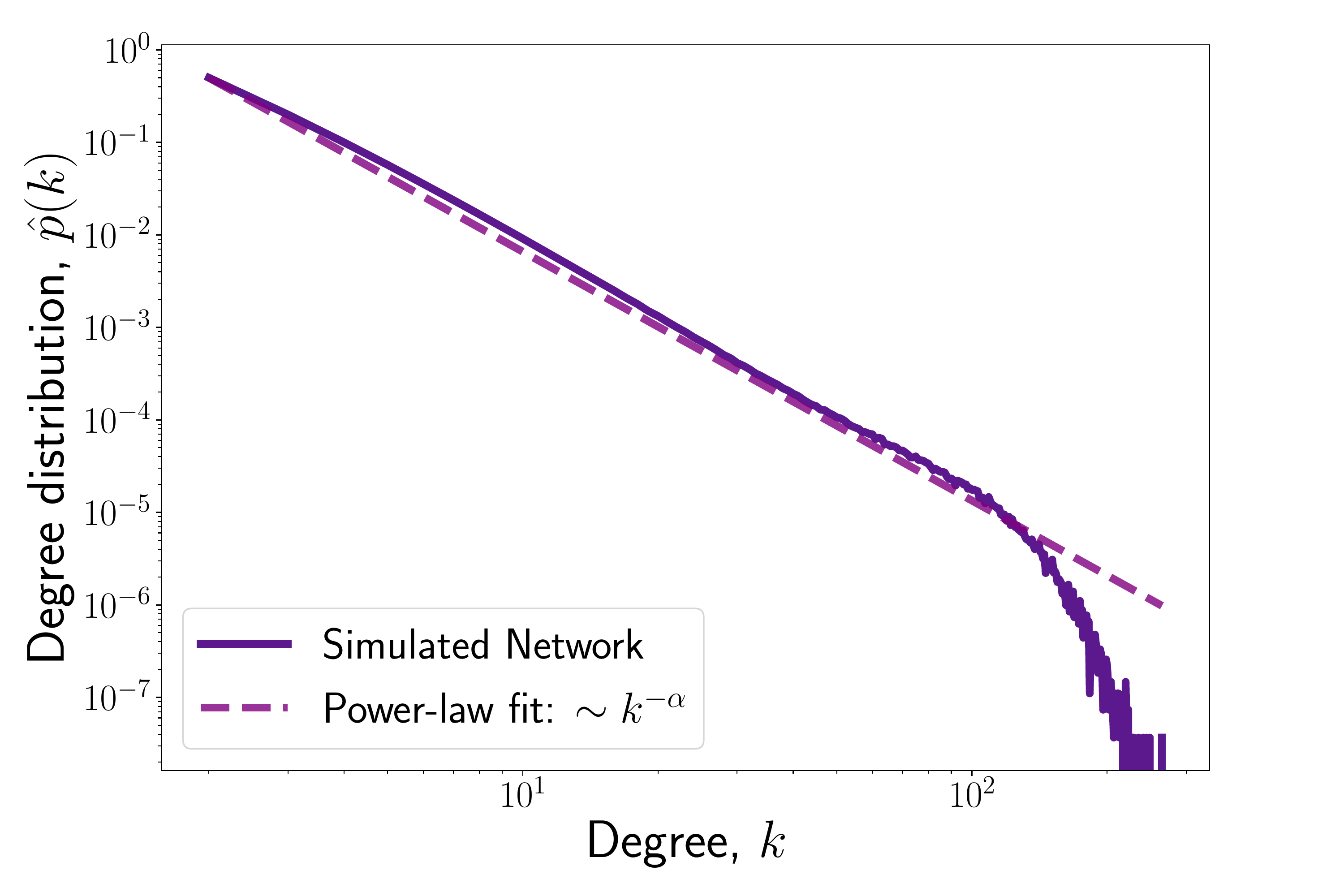}
\caption{\label{fig:degree_distrib}}
\end{subfigure}
\begin{subfigure}{.55\textwidth}
\centering
\includegraphics[width=\textwidth]{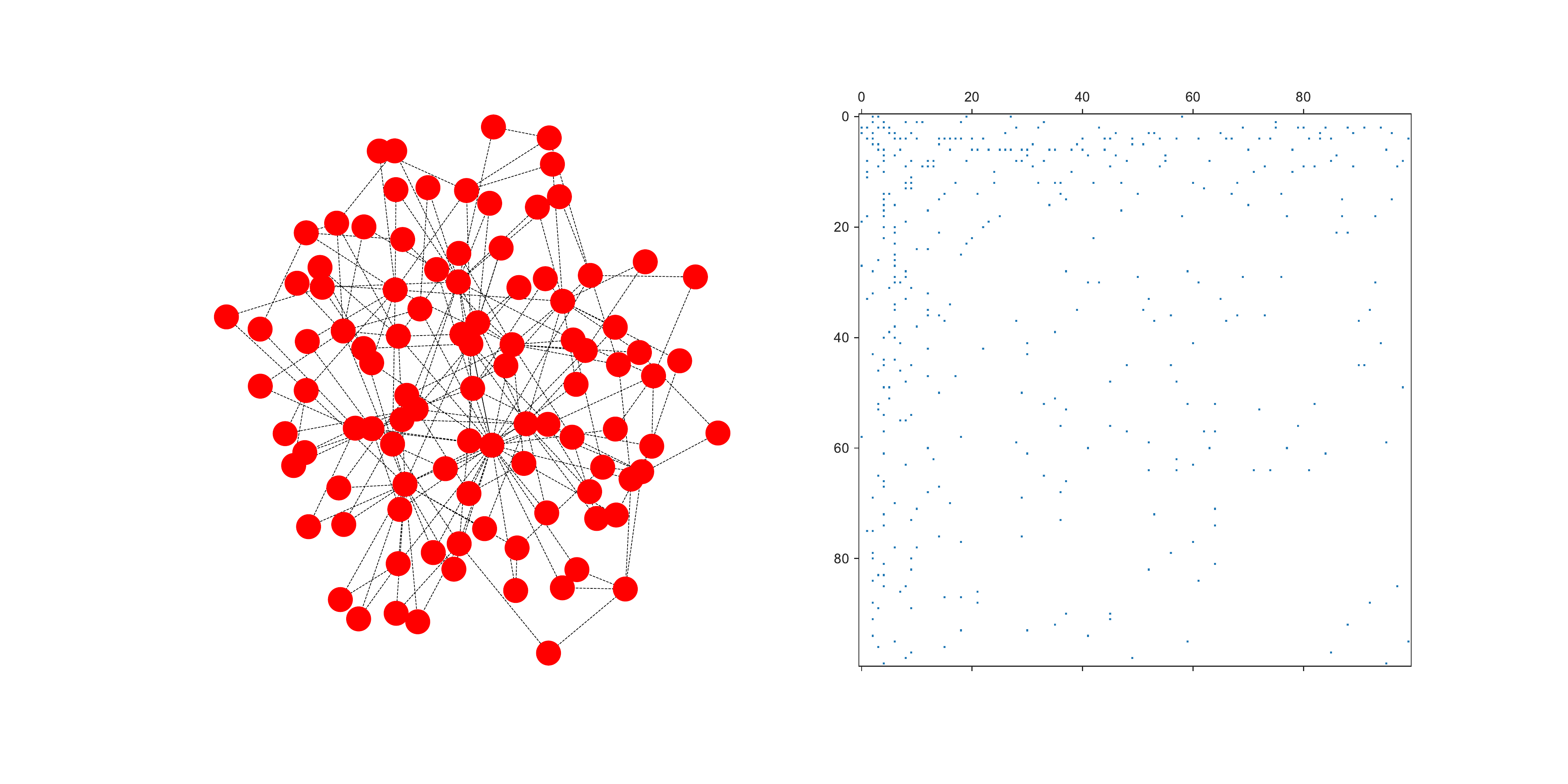}
\caption{\label{fig:network}}
\end{subfigure}
\caption{{\bf Scale-free network} {\bf (a)} Degree distribution of a random network generated by the preferential attachement {\it Barabasi-Albert} method. It is characterized by its power-law behavior. {\bf (b)} Visualization of a random network together with its adjacency matrix. N=100.\label{figs:network}}
\end{figure*}

The approximation $S\approx N$ is only valid when few individuals are infected and breaks down when the infection approaches its expected peak. From Eqs.~\ref{eq:ODE1}-~\ref{eq:ODE4}, it is possible to derive an expression for the number of infected individuals in the peak. Using the condition $\dot{I}=0$ and the constrain of constant total population size $N=S+E+I+R$, we have that
\begin{equation}
I_{\mathrm{peak}} \approx N\cdot \frac{\sigma\gamma}{\beta(\sigma+\gamma)}\left(\beta/\gamma - 1 - \log{\beta/\gamma}\right).
\end{equation}
We expect then deterministic exponential grow when $I(t)<I_{\mathrm{peak}}$. In Figure~\ref{fig:stochasticity1}, for the case of $\beta/\gamma=4.0$, the peak is reached at $I_{\mathrm{peak}}\approx1210$ individuals, and for that reason we see that the average epidemic trajectory grows exponentially for the whole range shown. On the other hand, in Figure~\ref{fig:stochasticity2} for the case $\beta/\gamma=1.3$, the peak is reached at  $I_{\mathrm{peak}}\approx 86$. Therefore, an exponential growth is observed only for a shorter period and the approximation made before is invalid for the later part.

\subsection*{Proliferation on a contact network}

We generate the random network of contacts by following a preferential attachment method~\cite{Barabasi1999}. Such a method produces graphs with heavy-tailed degree distribution $\hat p(k)\sim k^{-\alpha}$, like the one shown in Figure~\ref{fig:degree_distrib}. Under these circumstances, most of the individuals have few number of contacts close to the mean degree and a non-negligible amount of individuals have a large number of contacts. The degree of the highly connected individuals can reach values up to two orders of magnitude larger than the mean degree. In such case, a finite variance is guaranteed by a cutoff in the degree distribution at large degree values as a consequence of the finite size of the network.

In this context, a new variable $T_{nm}$ has been defined in the literature~\cite{Newman2002} to characterize the epidemiological dynamics on the random network, which accounts for the probability that the infection is transmitted through an edge in the network between nodes $n$ and $m$. The average value, $T$, is given in terms of the typical effective time of duration of an infection $\tau$, which is a exponentially distributed stochastic variable with mean value $\tau_\gamma = \gamma^{-1}$ , and the rate of infectious encounters between individuals in the network $\rho$. Given that in our model $\rho$ is proportional to $\beta$ (See Eq.~\ref{eq:StoE_N}), we derived that for the the case of a SEIR dynamics, $T$ is given by
\begin{eqnarray}
T &=& 1-\int_0^\infty \gamma e^{-\gamma\tau'}e^{-\frac{\beta}{\left\langle k \right\rangle}\tau'} d\tau' \nonumber\\
&=& 1-\frac{1}{1+\frac{\beta/\gamma}{\langle k\rangle}}.
\end{eqnarray}
where the second term correspond to the average probability across the network that no such transmission event occurs.

Moreover, a critical value $T_c=\left\langle k \right\rangle/(\left\langle k^2 \right\rangle - \left\langle k \right\rangle)$ has been derived for an uncorrelated network previously, which sets the minimum value of $T$ required for an outbreak to have the chance to turn into an epidemic. The value of $T_c$ corresponds to the correction factor in Eq.~\ref{eq:R0_N} that we derived for the renormalized basic reproductive number in the network of contacts.

As it has been derived before~\cite{Newman2002}, we start the derivation with a Dyson-like self-consistent equation for the probability $q_b$ that an individual at the end of a randomly selected interacting edge will not produce an epidemic, 
\begin{equation}
q_b=\frac{\sum_{k=1}^\infty  k\hat{p}(k) \Psi(T, k-1)}{\left\langle k \right\rangle},
\label{eq:u}
\end{equation}
where
\begin{equation}
\Psi(T, k) = \left(1-T+(Tq_b)\right)^{k}
\end{equation}
and $1-T$ is the probability that the disease does not spread through the interacting edge, $Tq_b$ is the probability that it spreads through the edge but it does not produce an epidemic and $\hat{p}_2(k)=k\hat{p}(k)/\left\langle k \right\rangle$ is the distribution of interacting edges connected to an individual at the end of a randomly selected interacting edge. This nearest-neighbors degree distribution is important because by following a randomly chosen interacting edge, is more likely to reach highly connected hubs than by randomly selecting single nodes. We solved Eq.~\ref{eq:u} numerically for a fixed value of $T$. Here, we are able to write the proliferation probability equations in terms of the epidemiological parameters that define the dynamics of Eqs.~\ref{eq:ODE1}-~\ref{eq:ODE4} and \ref{eq:StoE}-\ref{eq:ItoR} for the case of the SEIR model or by the epidemiological measurable values $\lambda$, $\tau_\gamma$ and $\tau_\sigma$.

Near the critical value $R_0 = 1$, some properties of the system show well defined scaling behavior \cite{PastorSatorras2015, Cohen2002}. In particular, the probability of having an epidemic $1-Q_{\rm ext} \sim(T-T_c)^{-\beta_{p}}$ with $\beta_c = 1/(3-\alpha)$ when $2<\alpha<3$, which is the case for the networks generated by the preferential attachment method and that corresponds to the most common networks occurring in nature~\cite{Cohen2002}. In our case, some approximations may become invalid due to large fluctuations. Given the relation that we derive between $T$ and $R_0$, it might be possible to derive new scaling laws for $R_0$ approaching its corresponding critical value. However, this is beyond the scope of this work.  

\subsection*{Random testing protocols and the likelihood function}

\begin{figure}[h!]
\includegraphics[width = \columnwidth]{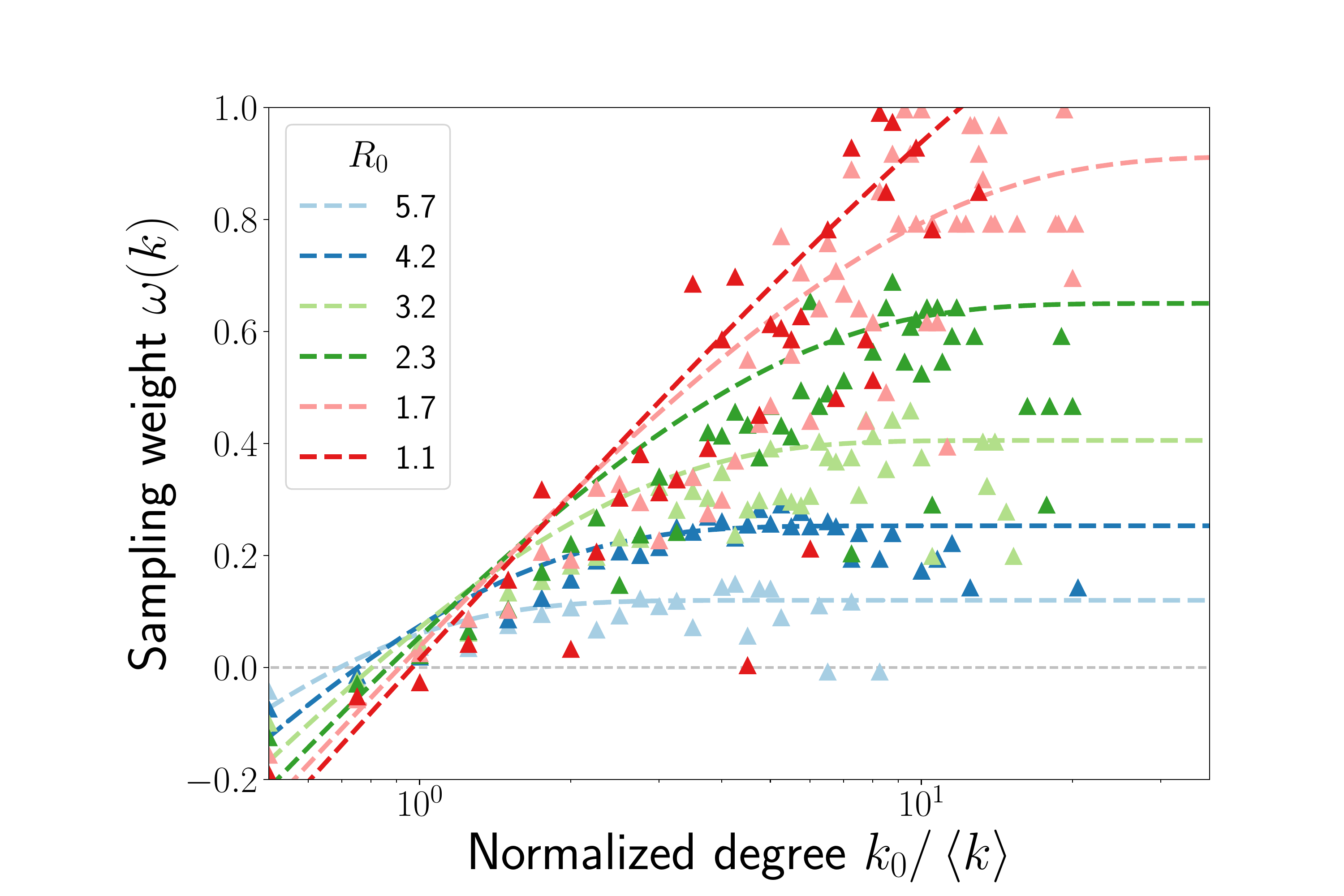}
\caption{{\bf Degree-based sampling weight} Log-ratio of probability that a degree $k$ is drawn from the network given that an outbreak seeded on it produced an epidemic and the degree distribution. Lines show analytical result and symbols simulation outcomes. Parameters $\gamma=1/6$, $\sigma=1/4$, $\alpha = 2.68$ \label{fig:loglikelihood}}
\end{figure}

The function $\omega(k)$ defined in the main text is proportional to the likelihood function $L(k)=p(\mathrm{proliferation}|k) = Q_p({R_0, k})$. It should be reduced to a constant value when the network of contacts is turned off because then $Q_p(R_0, k) =Q_p(R_0)  $

\begin{figure}[t!]
\centering
\begin{subfigure}{.98\columnwidth}
\includegraphics[width = \textwidth]{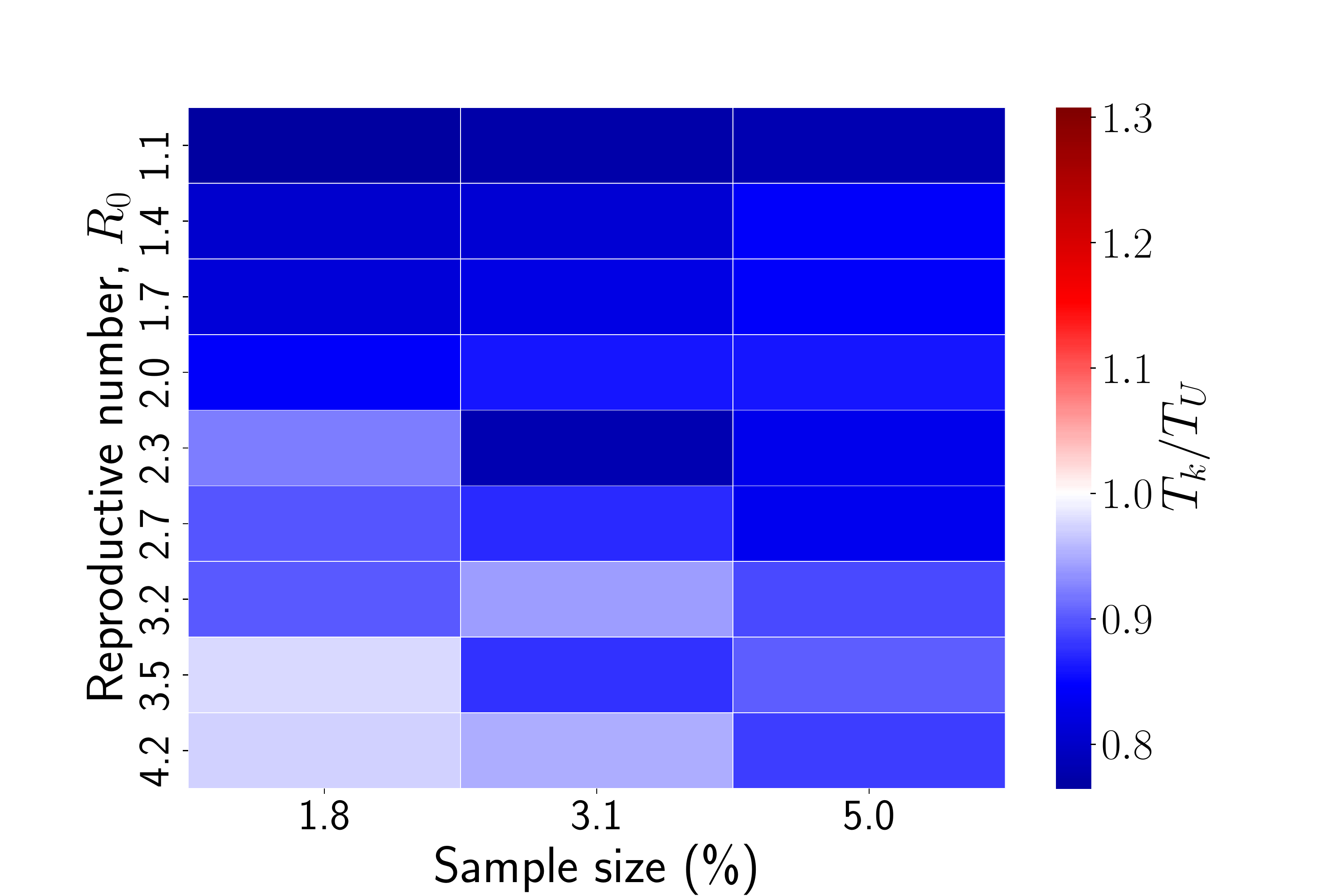}
\caption{}
\label{fig:sampling_t_N}
\end{subfigure}
\begin{subfigure}{.98\columnwidth}
\includegraphics[width = \textwidth]{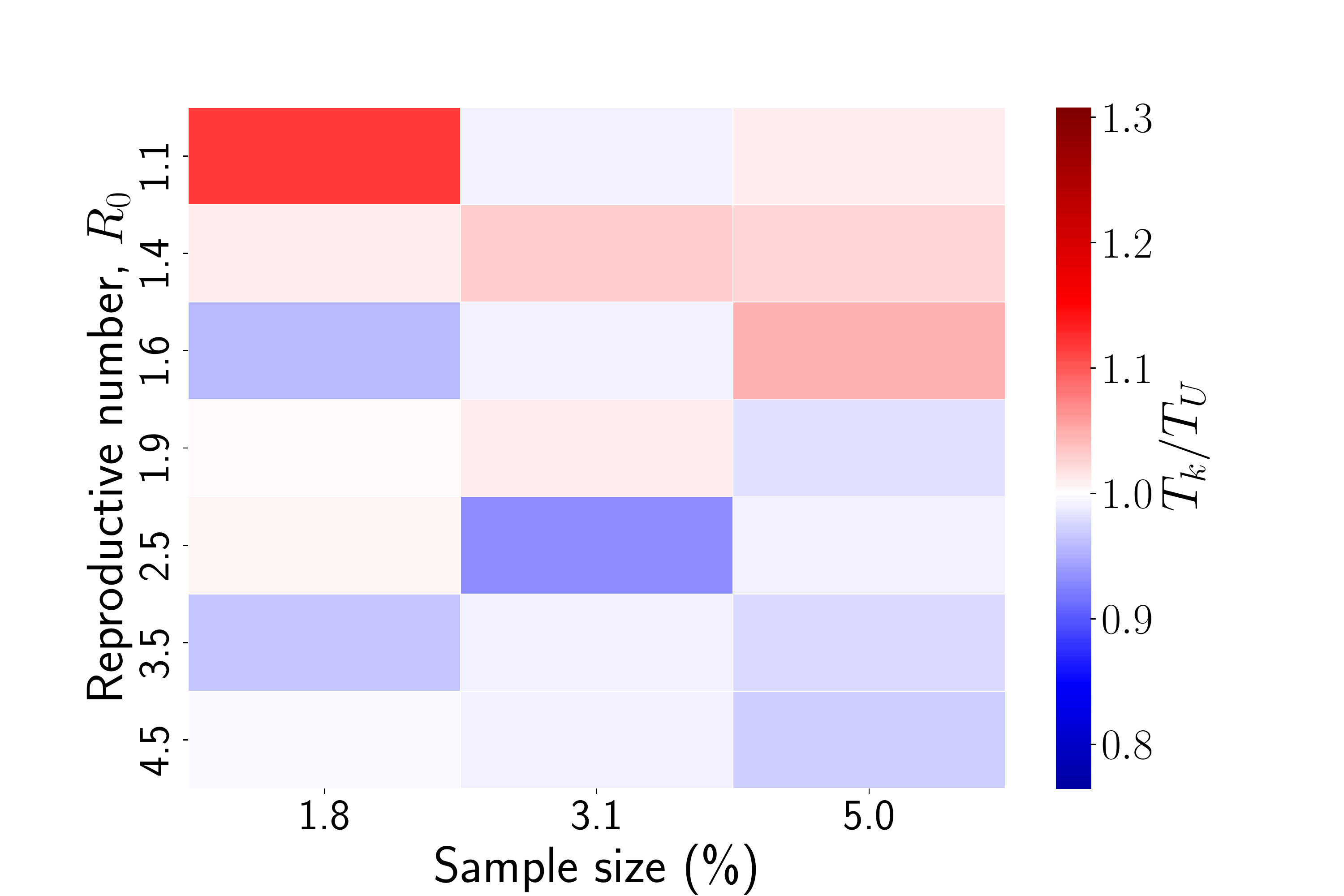}
\caption{}
\label{fig:sampling_t_FC}
\end{subfigure}
\caption{{\bf Degree-based random testing protocol.} Performance of the degree-based random testing protocol relative to the uniform protocol in terms of time of detection; we show the case of individuals interacting {\bf (a)} under a network of contacts and {\bf (b)} in a well-mixed population as a control.\label{figs:degree_based2}}
\end{figure}

Figure~\ref{fig:sampling_t_N} shows the outcomes of the simulations for the average time of first detection to complement the evaluation of the performance of the degree-based sampling testing protocol. Similar to the cluster size shown in Figure~\ref{fig:sampling_cs_N}, there is a significant improvement in the time at which the first infected individual is detected. For small values of $R_0$ the reduction of such time can be large than 20\%.

To test that the random testing protocol has an effect on a structured population characterized by a network of contacts, we simulate the degree-based sampling protocol turning off the network of interaction as a negative control but still using the likelihood function $L$ described above for the sampling of individuals. Figs~\ref{fig:sampling_cs_FC} and \ref{fig:sampling_t_FC} show the performance of the control protocol in terms of cluster size and time of the first detection, respectively.  In both cases the performance of the degree-based protocol is the same as the uniform sampling protocol.

\subsection*{Numerical Simulations}
The simulations were performed with a modified version of the individual-based SEIR python code \verb|seirsplus| publicly available in \url{https://github.com/ryansmcgee/seirsplus}. The simulations consist of a Gillespie algorithm for independent Poisson processes.


\printbibliography[title=References]

@article{Altarelli2014,
  title = {Bayesian Inference of Epidemics on Networks via Belief Propagation},
  author = {Altarelli, Fabrizio and Braunstein, Alfredo and Dall'Asta, Luca and Lage-Castellanos, Alejandro and Zecchina, Riccardo},
  journal = {Phys. Rev. Lett.},
  volume = {112},
  issue = {11},
  pages = {118701},
  numpages = {5},
  year = {2014},
  month = {Mar},
  publisher = {American Physical Society},
  doi = {10.1103/PhysRevLett.112.118701},
}

@article{Lokhov2014,
  title = {Inferring the origin of an epidemic with a dynamic message-passing algorithm},
  author = {Lokhov, Andrey Y. and M\'ezard, Marc and Ohta, Hiroki and Zdeborov\'a, Lenka},
  journal = {Phys. Rev. E},
  volume = {90},
  issue = {1},
  pages = {012801},
  numpages = {9},
  year = {2014},
  month = {Jul},
  publisher = {American Physical Society},
  doi = {10.1103/PhysRevE.90.012801},
}

@article{Lokhov2015,
  title = {Dynamic message-passing equations for models with unidirectional dynamics},
  author = {Lokhov, Andrey Y. and M\'ezard, Marc and Zdeborov\'a, Lenka},
  journal = {Phys. Rev. E},
  volume = {91},
  issue = {1},
  pages = {012811},
  numpages = {22},
  year = {2015},
  month = {Jan},
  publisher = {American Physical Society},
  doi = {10.1103/PhysRevE.91.012811},
}

@book{Bailey1975,
  title={The Mathematical Theory of Infectious Diseases and Its Applications},
  author={Bailey, N.T.J.},
  isbn={9780852642313},
  lccn={76675653},
  series={Mathematics in Medicine Series},
  year={1975},
  publisher={Griffin}
}

@article{Allen2017,
author = {Allen, Linda J.S.},
doi = {10.1016/j.idm.2017.03.001},
issn = {24680427},
journal = {Infectious Disease Modelling},
number = {2},
pages = {128--142},
publisher = {Elsevier Ltd},
title = {{A primer on stochastic epidemic models: Formulation, numerical simulation, and analysis}},
volume = {2},
year = {2017}
}

@article{Tritch2018,
author = {Tritch, William and Allen, Linda J.S.},
doi = {10.1016/j.idm.2018.03.002},
issn = {24680427},
journal = {Infectious Disease Modelling},
pages = {60--73},
publisher = {Elsevier Ltd},
title = {{Duration of a minor epidemic}},
volume = {3},
year = {2018}
}

@article{Newman2002,
  title = {Spread of epidemic disease on networks},
  author = {Newman, M. E. J.},
  journal = {Phys. Rev. E},
  volume = {66},
  issue = {1},
  pages = {016128},
  numpages = {11},
  year = {2002},
  month = {Jul},
  publisher = {American Physical Society}
}

@article{Meyers2005,
title = {Network theory and SARS: predicting outbreak diversity},
journal = {Journal of Theoretical Biology},
volume = {232},
number = {1},
pages = {71-81},
year = {2005},
issn = {0022-5193},
author = {Lauren Ancel Meyers and Babak Pourbohloul and M.E.J. Newman and Danuta M. Skowronski and Robert C. Brunham}
}

@Article{Eubank2004,
author={Eubank, Stephen and Guclu, Hasan and Anil Kumar, V. S. and Marathe, Madhav V. and Srinivasan, Aravind and Toroczkai, Zolt{\'a}n
and Wang, Nan},
title={Modelling disease outbreaks in realistic urban social networks},
journal={Nature},
year={2004},
month={May},
day={01},
volume={429},
number={6988},
pages={180-184},
issn={1476-4687}
}

@Article{Lloyd-Smith2005,
author={Lloyd-Smith, J. O. and Schreiber, S. J. and Kopp, P. E. and Getz, W. M.},
title={Superspreading and the effect of individual variation on disease emergence},
journal={Nature},
year={2005},
month={Nov},
day={01},
volume={438},
number={7066},
pages={355-359},
issn={1476-4687}
}

@misc{Allard2020,
      title={The role of directionality, heterogeneity and correlations in epidemic risk and spread}, 
      author={Antoine Allard and Cristopher Moore and Samuel V. Scarpino and Benjamin M. Althouse and Laurent Hébert-Dufresne},
      year={2020},
      archivePrefix={arXiv},
      primaryClass={physics.soc-ph}
}

@article{Althouse2020,
    author = {Althouse, Benjamin M. AND Wenger, Edward A. AND Miller, Joel C. AND Scarpino, Samuel V. AND Allard, Antoine AND Hébert-Dufresne, Laurent AND Hu, Hao},
    journal = {PLOS Biology},
    publisher = {Public Library of Science},
    title = {Superspreading events in the transmission dynamics of SARS-CoV-2: Opportunities for interventions and control},
    year = {2020},
    month = {11},
    volume = {18},
    pages = {1-13},
    number = {11}
}

@article {Reich2020,
	author = {Reich, Ofir and Shalev, Guy and Kalvari, Tom},
	title = {Modeling COVID-19 on a network: super-spreaders, testing and containment},
	elocation-id = {2020.04.30.20081828},
	year = {2020},
	doi = {10.1101/2020.04.30.20081828},
	publisher = {Cold Spring Harbor Laboratory Press},
	URL = {https://www.medrxiv.org/content/early/2020/05/05/2020.04.30.20081828},
	journal = {medRxiv}
}

@article{Cowling2020,
author = {Cowling, Benjamin J. and Ali, Sheikh Taslim and Ng, Tiffany W.Y. and Tsang, Tim K. and Li, Julian C.M. and Fong, Min Whui and Liao, Qiuyan and Kwan, Mike YW and Lee, So Lun and Chiu, Susan S. and Wu, Joseph T. and Wu, Peng and Leung, Gabriel M.},
doi = {10.1016/S2468-2667(20)30090-6},
issn = {24682667},
journal = {The Lancet Public Health},
number = {5},
pages = {e279--e288},
pmid = {32311320},
publisher = {The Author(s). Published by Elsevier Ltd. This is an Open Access article under the CC BY 4.0 license},
title = {{Impact assessment of non-pharmaceutical interventions against coronavirus disease 2019 and influenza in Hong Kong: an observational study}},
volume = {5},
year = {2020}
}

@Article{Miller2020,
author={Miller, Danielle and Martin, Michael A. and Harel, Noam and Tirosh, Omer and Kustin, Talia and Meir, Moran and Sorek, Nadav and Gefen-Halevi, Shiraz and Amit, Sharon and Vorontsov, Olesya and Shaag, Avraham and Wolf, Dana and Peretz, Avi and Shemer-Avni, Yonat and Roif-Kaminsky, Diana and Kopelman, Naama M. and Huppert, Amit and Koelle, Katia and Stern, Adi},
title={Full genome viral sequences inform patterns of SARS-CoV-2 spread into and within Israel},
journal={Nature Communications},
year={2020},
month={Nov},
day={02},
volume={11},
number={1},
pages={5518},
issn={2041-1723}
}

@article{Li2020,
author = {Ruiyun Li  and Sen Pei  and Bin Chen  and Yimeng Song  and Tao Zhang  and Wan Yang  and Jeffrey Shaman },
title = {Substantial undocumented infection facilitates the rapid dissemination of novel coronavirus (SARS-CoV-2)},
journal = {Science},
volume = {368},
number = {6490},
pages = {489-493},
year = {2020}
}

@Article{Firth2020,
author={Firth, Josh A. and Hellewell, Joel and Klepac, Petra and Kissler, Stephen and Jit, Mark and Atkins, Katherine E. and Clifford, Samuel and Villabona-Arenas, C. Julian and Meakin, Sophie R. and Diamond, Charlie and Bosse, Nikos I. and Munday, James D. and Prem, Kiesha and Foss, Anna M. and Nightingale, Emily S. and Zandvoort, Kevin van and Davies, Nicholas G. and Gibbs, Hamish P. and Medley, Graham and Gimma, Amy and Flasche, Stefan and Simons, David and Auzenbergs, Megan and Russell, Timothy W. and Quilty, Billy J. and Rees, Eleanor M. and Leclerc, Quentin J. and Edmunds, W. John and Funk, Sebastian and Houben, Rein M. G. J. and Knight, Gwenan M. and Abbott, Sam and Sun, Fiona Yueqian and Lowe, Rachel and Tully, Damien C. and Procter, Simon R. and Jarvis, Christopher I. and Endo, Akiram and O'Reilly, Kathleenm and Emery, Jon C. and Jombart, Thibautmand Rosello, Aliciam and Deol, Arminder K.mand Quaife, Matthewm and Hu{\'e}, St{\'e}phanemand Liu, Yang and Eggo, Rosalind M. and Pearson, Carl A. B. and Kucharski, Adam J. and Spurgin, Lewis G. and Group, CMMID COVID-19 Working},
title={Using a real-world network to model localized COVID-19 control strategies},
journal={Nature Medicine},
year={2020},
month={Oct},
day={01},
volume={26},
number={10},
pages={1616-1622},
issn={1546-170X}
}

@article {Rivett2020,
article_type = {journal},
title = {Screening of healthcare workers for SARS-CoV-2 highlights the role of asymptomatic carriage in COVID-19 transmission},
author = {Rivett, Lucy and Sridhar, Sushmita and Sparkes, Dominic and Routledge, Matthew and Jones, Nick K and Forrest, Sally and Young, Jamie and Pereira-Dias, Joana and Hamilton, William L and Ferris, Mark and Torok, M Estee and Meredith, Luke and The CITIID-NIHR COVID-19 BioResource Collaboration and Curran, Martin D and Fuller, Stewart and Chaudhry, Afzal and Shaw, Ashley and Samworth, Richard J and Bradley, John R and Dougan, Gordon and Smith, Kenneth GC and Lehner, Paul J and Matheson, Nicholas J and Wright, Giles and Goodfellow, Ian G and Baker, Stephen and Weekes, Michael P},
volume = 9,
year = 2020,
month = {may},
pub_date = {2020-05-11},
pages = {e58728},
citation = {eLife 2020;9:e58728},
journal = {eLife},
issn = {2050-084X},
publisher = {eLife Sciences Publications, Ltd},
}

@article{Muller2020,
   title={Testing of asymptomatic individuals for fast feedback-control of COVID-19 pandemic},
   volume={17},
   ISSN={1478-3975},
   number={6},
   journal={Physical Biology},
   publisher={IOP Publishing},
   author={Müller, Markus and Derlet, Peter M and Mudry, Christopher and Aeppli, Gabriel},
   year={2020},
   month={Oct},
   pages={065007}
}

@article {Karin2020,
	author = {Karin, Omer and Bar-On, Yinon M. and Milo, Tomer and Katzir, Itay and Mayo, Avi and Korem, Yael and Dudovich, Boaz and Yashiv, Eran and Zehavi, Amos J. and Davidovitch, Nadav and Milo, Ron and Alon, Uri},
	title = {Cyclic exit strategies to suppress COVID-19 and allow economic activity},
	elocation-id = {2020.04.04.20053579},
	year = {2020},
	doi = {10.1101/2020.04.04.20053579},
	publisher = {Cold Spring Harbor Laboratory Press},
	URL = {https://www.medrxiv.org/content/early/2020/04/28/2020.04.04.20053579},
	journal = {medRxiv}
}

@article{Walker2020,
author = {Patrick G. T. Walker  and Charles Whittaker  and Oliver J. Watson  and Marc Baguelin  and Peter Winskill  and Arran Hamlet  and Bimandra A. Djafaara  and Zulma Cucunubá  and Daniela Olivera Mesa  and Will Green  and Hayley Thompson  and Shevanthi Nayagam  and Kylie E. C. Ainslie  and Sangeeta Bhatia  and Samir Bhatt  and Adhiratha Boonyasiri  and Olivia Boyd  and Nicholas F. Brazeau  and Lorenzo Cattarino  and Gina Cuomo-Dannenburg  and Amy Dighe  and Christl A. Donnelly  and Ilaria Dorigatti  and Sabine L. van Elsland  and Rich FitzJohn  and Han Fu  and Katy A. M. Gaythorpe  and Lily Geidelberg  and Nicholas Grassly  and David Haw  and Sarah Hayes  and Wes Hinsley  and Natsuko Imai  and David Jorgensen  and Edward Knock  and Daniel Laydon  and Swapnil Mishra  and Gemma Nedjati-Gilani  and Lucy C. Okell  and H. Juliette Unwin  and Robert Verity  and Michaela Vollmer  and Caroline E. Walters  and Haowei Wang  and Yuanrong Wang  and Xiaoyue Xi  and David G. Lalloo  and Neil M. Ferguson  and Azra C. Ghani },
title = {The impact of COVID-19 and strategies for mitigation and suppression in low- and middle-income countries},
journal = {Science},
volume = {369},
number = {6502},
pages = {413-422},
year = {2020}
}

@article {BarOn2020,
article_type = {journal},
title = {Science Forum: SARS-CoV-2 (COVID-19) by the numbers},
author = {Bar-On, Yinon M and Flamholz, Avi and Phillips, Rob and Milo, Ron},
volume = 9,
year = 2020,
month = {mar},
pub_date = {2020-03-31},
pages = {e57309},
citation = {eLife 2020;9:e57309},
journal = {eLife},
issn = {2050-084X},
publisher = {eLife Sciences Publications, Ltd}
}

@article{Flaxman2020,
author = {Flaxman, Seth and Mishra, Swapnil and Gandy, Axel and Unwin, H. Juliette T. and Mellan, Thomas A. and Coupland, Helen and Whittaker, Charles and Zhu, Harrison and Berah, Tresnia and Eaton, Jeffrey W. and Monod, M{\'{e}}lodie and Perez-Guzman, Pablo N. and Schmit, Nora and Cilloni, Lucia and Ainslie, Kylie E.C. and Baguelin, Marc and Boonyasiri, Adhiratha and Boyd, Olivia and Cattarino, Lorenzo and Cooper, Laura V. and Cucunub{\'{a}}, Zulma and Cuomo-Dannenburg, Gina and Dighe, Amy and Djaafara, Bimandra and Dorigatti, Ilaria and van Elsland, Sabine L. and FitzJohn, Richard G. and Gaythorpe, Katy A.M. and Geidelberg, Lily and Grassly, Nicholas C. and Green, William D. and Hallett, Timothy and Hamlet, Arran and Hinsley, Wes and Jeffrey, Ben and Knock, Edward and Laydon, Daniel J. and Nedjati-Gilani, Gemma and Nouvellet, Pierre and Parag, Kris V. and Siveroni, Igor and Thompson, Hayley A. and Verity, Robert and Volz, Erik and Walters, Caroline E. and Wang, Haowei and Wang, Yuanrong and Watson, Oliver J. and Winskill, Peter and Xi, Xiaoyue and Walker, Patrick G.T. and Ghani, Azra C. and Donnelly, Christl A. and Riley, Steven and Vollmer, Michaela A.C. and Ferguson, Neil M. and Okell, Lucy C. and Bhatt, Samir},
doi = {10.1038/s41586-020-2405-7},
isbn = {4158602024057},
issn = {14764687},
journal = {Nature},
number = {7820},
pages = {257--261},
pmid = {32512579},
title = {{Estimating the effects of non-pharmaceutical interventions on COVID-19 in Europe}},
volume = {584},
year = {2020}
}

@Article{He2020,
author={He, Xi and Lau, Eric H. Y. and Wu, Peng and Deng, Xilong and Wang, Jian and Hao, Xinxin and Lau, Yiu Chung and Wong, Jessica Y. and Guan, Yujuan and Tan, Xinghua and Mo, Xiaoneng and Chen, Yanqing and Liao, Baolin and Chen, Weilie and Hu, Fengyu and Zhang, Qing and Zhong, Mingqiu and Wu, Yanrong and Zhao, Lingzhai and Zhang, Fuchun and Cowling, Benjamin J. and Li, Fang and Leung, Gabriel M.},
title={Temporal dynamics in viral shedding and transmissibility of COVID-19},
journal={Nature Medicine},
year={2020},
month={May},
day={01},
volume={26},
number={5},
pages={672-675},
issn={1546-170X}
}

@Article{Black2020,
author={Black, James R. M. and Bailey, Chris and Przewrocka, Joanna and Dijkstra, Krijn K. and Swanton, Charles},
title={COVID-19: the case for health-care worker screening to prevent hospital transmission},
journal={The Lancet},
year={2020},
month={May},
day={02},
publisher={Elsevier},
volume={395},
number={10234},
pages={1418-1420},
issn={0140-6736}
}

@article {Goyal2021,
article_type = {journal},
title = {Viral load and contact heterogeneity predict SARS-CoV-2 transmission and super-spreading events},
author = {Goyal, Ashish and Reeves, Daniel B and Cardozo-Ojeda, E Fabian and Schiffer, Joshua T and Mayer, Bryan T},
editor = {Walczak, Aleksandra M and Childs, Lauren and Forde, Jonathan},
volume = 10,
year = 2021,
month = {feb},
pub_date = {2021-02-23},
pages = {e63537},
citation = {eLife 2021;10:e63537},
journal = {eLife},
issn = {2050-084X},
publisher = {eLife Sciences Publications, Ltd}
}

@article{Jones2021,
author = {Terry C. Jones  and Guido Biele  and Barbara Mühlemann  and Talitha Veith  and Julia Schneider  and Jörn Beheim-Schwarzbach  and Tobias Bleicker  and Julia Tesch  and Marie Luisa Schmidt  and Leif Erik Sander  and Florian Kurth  and Peter Menzel  and Rolf Schwarzer  and Marta Zuchowski  and Jörg Hofmann  and Andi Krumbholz  and Angela Stein  and Anke Edelmann  and Victor Max Corman  and Christian Drosten },
title = {Estimating infectiousness throughout SARS-CoV-2 infection course},
journal = {Science},
volume = {373},
number = {6551},
pages = {eabi5273},
year = {2021}
}

@article {Qiu2021,
	author = {Qiu, Xueting and Miller, Joel C and MacFadden, Derek R and Hanage, William P},
	title = {Evaluating the contributions of strategies to prevent SARS-CoV-2 transmission in the healthcare setting: a modelling study},
	volume = {11},
	number = {3},
	elocation-id = {e044644},
	year = {2021},
	publisher = {British Medical Journal Publishing Group},
	issn = {2044-6055},
	journal = {BMJ Open}
}

@article{Barabasi1999,
author = {Albert-László Barabási  and Réka Albert },
title = {Emergence of Scaling in Random Networks},
journal = {Science},
volume = {286},
number = {5439},
pages = {509-512},
year = {1999},
publisher = {American Association for the Advancement of Science}
}

@article{McCormick2010	,
author = {Tyler H.  McCormick and Matthew J. Salganik and Tian Zheng},
title = {How Many People Do You Know?: Efficiently Estimating Personal Network Size},
journal = {Journal of the American Statistical Association},
volume = {105},
number = {489},
pages = {59-70},
year  = {2010},
publisher = {Taylor & Francis}
}

@article{Killworth1990,
title = {Estimating the size of personal networks},
journal = {Social Networks},
volume = {12},
number = {4},
pages = {289-312},
year = {1990},
issn = {0378-8733},
author = {Peter D. Killworth and Eugene C. Johnsen and H.Russell Bernard and Gene {Ann Shelley} and Christopher McCarty}
}

@article{Maltiel2015,
author = {Rachael Maltiel and Adrian E. Raftery and Tyler H. McCormick and Aaron J. Baraff},
title = {{Estimating population size using the network scale up method}},
volume = {9},
journal = {The Annals of Applied Statistics},
number = {3},
publisher = {Institute of Mathematical Statistics},
pages = {1247 -- 1277},
year = {2015}
}

@article{Russell1991,
title = {Estimating the size of an average personal network and of an event subpopulation: Some empirical results},
journal = {Social Science Research},
volume = {20},
number = {2},
pages = {109-121},
year = {1991},
issn = {0049-089X},
author = {H {Russell Bernard} and Eugene C Johnsen and Peter D Killworth and Scott Robinson}
}

@article{Mutesa2021,
author = {Mutesa, Leon and Ndishimye, Pacifique and Butera, Yvan and Souopgui, Jacob and Uwineza, Annette and Rutayisire, Robert and Ndoricimpaye, Ella Larissa and Musoni, Emile and Rujeni, Nadine and Nyatanyi, Thierry and Ntagwabira, Edouard and Semakula, Muhammed and Musanabaganwa, Clarisse and Nyamwasa, Daniel and Ndashimye, Maurice and Ujeneza, Eva and Mwikarago, Ivan Emile and Muvunyi, Claude Mambo and Mazarati, Jean Baptiste and Nsanzimana, Sabin and Turok, Neil and Ndifon, Wilfred},
issn = {14764687},
journal = {Nature},
number = {7841},
pages = {276--280},
pmid = {33086375},
publisher = {Springer US},
title = {{A pooled testing strategy for identifying SARS-CoV-2 at low prevalence}},
volume = {589},
year = {2021}
}

@article{Baker2021,
author = {Antoine Baker  and Indaco Biazzo  and Alfredo Braunstein  and Giovanni Catania  and Luca Dall’Asta  and Alessandro Ingrosso  and Florent Krzakala  and Fabio Mazza  and Marc Mézard  and Anna Paola Muntoni  and Maria Refinetti  and Stefano Sarao Mannelli  and Lenka Zdeborová },
title = {Epidemic mitigation by statistical inference from contact tracing data},
journal = {Proceedings of the National Academy of Sciences},
volume = {118},
number = {32},
pages = {e2106548118},
year = {2021},
doi = {10.1073/pnas.2106548118},
}

@article{Yuan2021,
title = {The Prediction for COVID-19 Outbreak in China by using the Concept of Term Structure for the Turning Period},
journal = {Procedia Computer Science},
volume = {187},
pages = {284-293},
year = {2021},
note = {2020 International Conference on Identification, Information and Knowledge in the Internet of Things, IIKI2020},
issn = {1877-0509},
author = {George X. Yuan and Lan Di and Zheng Yang and Guoqi Qian and Xiaosong Qian and Tu Zeng},
}

@book{keeling2008,
  title={Modeling Infectious Diseases in Humans and Animals},
  author={Keeling, M.J. and Rohani, P.},
  isbn={9780691116174},
  lccn={2006939548},
  year={2008},
  publisher={Princeton University Press}
}

@article{PastorSatorras2015,
  title = {Epidemic processes in complex networks},
  author = {Pastor-Satorras, Romualdo and Castellano, Claudio and Van Mieghem, Piet and Vespignani, Alessandro},
  journal = {Rev. Mod. Phys.},
  volume = {87},
  issue = {3},
  pages = {925--979},
  numpages = {55},
  year = {2015},
  month = {Aug},
  publisher = {American Physical Society},
}

@article{Cohen2002,
  title = {Percolation critical exponents in scale-free networks},
  author = {Cohen, Reuven and ben-Avraham, Daniel and Havlin, Shlomo},
  journal = {Phys. Rev. E},
  volume = {66},
  issue = {3},
  pages = {036113},
  numpages = {4},
  year = {2002},
  month = {Sep},
  publisher = {American Physical Society},
  doi = {10.1103/PhysRevE.66.036113},
}

@article{small2005,
author = {Small,  Michael and Tse, Chi K.},
title = {SMALL WORLD AND SCALE FREE MODEL OF TRANSMISSION OF SARS},
journal = {International Journal of Bifurcation and Chaos},
volume = {15},
number = {05},
pages = {1745-1755},
year = {2005},
doi = {10.1142/S0218127405012776},
}

@article{Allen2012,
author = { Linda J. S.   Allen  and  Glenn   E.   Lahodny   Jr },
title = {Extinction thresholds in deterministic and stochastic epidemic models},
journal = {Journal of Biological Dynamics},
volume = {6},
number = {2},
pages = {590-611},
year  = {2012},
publisher = {Taylor & Francis},
doi = {10.1080/17513758.2012.665502},
}

\end{document}